\documentclass[pra,twoside,twocolumn,superscriptaddress,floatfix,APS,10pt]{revtex4}
\usepackage{eurosym}
\usepackage[utf8]{inputenc}
\usepackage{amsmath}
\usepackage{amsfonts}
\usepackage{t4phonet}
\usepackage{amssymb}
\usepackage{setspace}
\usepackage{tipa}
\usepackage{epsfig}
\usepackage{graphicx}
\usepackage{float}
\usepackage{times}
\usepackage{subfigure}
\usepackage{hyperref}
\hypersetup{colorlinks=true, urlcolor=blue, linkcolor=blue, citecolor=blue}
\usepackage[pdftex,dvipsnames]{xcolor}
\usepackage[colorinlistoftodos,prependcaption,textsize=normalsize]{todonotes}
\usepackage{epstopdf}
\usepackage[a4paper]{geometry}
\pdfoutput=1

\DeclareMathSizes{10}{9}{7}{6}

\usepackage{newtxtext,newtxmath}

\usepackage{fancyhdr}
\fancypagestyle{title}{
  \setlength{\headheight}{0pt}%
  \fancyhf{}
  \fancyfoot[L]{\footnotesize 2469-9926/2018/98(6)/063841(11)}
  \fancyfoot[C]{\footnotesize 063841-\thepage}
  \fancyfoot[R]{\footnotesize \copyright2018 American Physical Society}
  \fancyhead[C]{PHYSICAL REVIEW A {\bf 98}, 063841 (2018)}
}

\begin{document}
\vspace*{0.3cm}
\title{\normalsize Solitons in a chain of charge-parity-symmetric dimers}
\author{\small O. B. Kirikchi}
\affiliation{\footnotesize Department of Mathematical Sciences, University of Essex, Wivenhoe Park, Colchester CO4 3SQ, United Kingdom}
\author{\small Boris A. Malomed}
\affiliation{\footnotesize Department of Physical Electronics, School of Electrical Engineering, Faculty of Engineering, and the Center for Light-Matter Interaction, Tel Aviv University, 69978 Tel Aviv, Israel}
\author{\small N. Karjanto}
\email{natanael@skku.edu}
\affiliation{\footnotesize Department of Mathematics, University College, Sungkyunkwan University, Natural Science Campus, 2066 Seobu-ro, Jangan-gu, Suwon 16419, Gyeonggi-do, Republic of Korea}
\author{\small R. Kusdiantara}
\affiliation{\footnotesize Department of Mathematical Sciences, University of Essex, Wivenhoe Park, Colchester CO4 3SQ, United Kingdom}
\affiliation{\footnotesize Center for Research on Mathematical Modeling and Simulation, Bandung Institute of Technology, Labtek III, First Floor, Jalan Ganesha 10, Bandung 40132, Indonesia}
\author{\small H. Susanto}
\affiliation{\footnotesize Department of Mathematical Sciences, University of Essex, Wivenhoe Park, Colchester CO4 3SQ, United Kingdom}

\begin{abstract}
\begin{center}
{\footnotesize \href{http://crossmark.crossref.org/dialog/?doi=10.1103/PhysRevA.98.063841&domain=pdf&date_stamp=2018-12-26}{\includegraphics[scale=0.04]{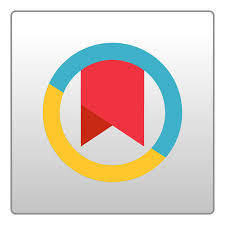}} \ (Received 28 June 2018; published 26 December 2018)}
\end{center}

We consider an array of dual-core waveguides, which represent an optical realization of a chain of dimers, with an active (gain-loss) coupling between the cores, opposite signs of discrete diffraction in the parallel arrays, and a phase-velocity mismatch between them (which is necessary for the stability of the system). The array provides an optical emulation of the charge-parity ($\mathcal{CP}$) symmetry. The addition of the intracore cubic nonlinearity gives rise to several species of fundamental discrete solitons, which exist in continuous families, although the system is non-Hermitian. The existence and stability of the soliton families are explored by means of analytical and numerical methods. An asymptotic analysis is presented for the case of weak intersite coupling (i.e., near the anticontinuum limit), as well as weak coupling between cores in each dimer. Several families of fundamental discrete solitons are found in the semi-infinite gap of the system's spectrum, that have no counterparts in the continuum limit, as well as a branch which belongs to the finite band gap and carries over into a family of stable gap solitons in that limit. One branch develops an oscillatory instability above a critical strength of the intersite coupling, others being stable in their entire existence regions. Unlike solitons in conservative lattices, which are controlled solely by the strength of the intersite coupling, here fundamental-soliton families have several control parameters, one of which, viz., the coefficient of the intercore coupling in the active host medium, may be readily adjusted in the experiment by varying the gain applied to the medium.\\

\noindent
DOI: \href{https://doi.org/10.1103/PhysRevA.98.063841}{10.1103/PhysRevA.98.063841}
\end{abstract}
\maketitle
\thispagestyle{title}

\section{Introduction}

Charge-parity ($\mathcal{CP}$) symmetry is one of the fundamental principles in physics of elementary particles~\cite{QFT}. Except for the small violation by weak nuclear forces, it holds for all interactions~\cite{weak}. The $\mathcal{CP}$ operator is the product of the parity transformation, $\mathcal{P}$, which reverses the coordinates, and charge conjugation, $\mathcal{C}$, which interchanges particles and antiparticles, i.e., essentially, positive and negative electric charges.

While the usual derivation of the $\mathcal{CP}$ symmetry is performed for Hermitian Hamiltonians, this symmetry may hold for Hamiltonians which are not Hermitian~\cite{Ham}. In fact, Hamiltonians which commute with another symmetry operator, viz., the parity-time one, $\mathcal{PT}$ ($\mathcal{T}$ is the time-inverting transform), may include an anti-Hermitian spatially antisymmetric (odd) part, provided that the Hermitian one has a spatially even structure~\cite{Bender}. The spectrum of energy eigenvalues, generated by such $\mathcal{PT}$-symmetric non-Hermitian Hamiltonians, may be purely real (i.e., physically relevant) up to a critical strength of the anti-Hermitian term, at which the $\mathcal{PT}$ symmetry is broken, making the system (in most cases) physically irrelevant above this point.

It is well known that non-Hermitian $\mathcal{PT}$-symmetric Hamiltonians may be emulated theoretically~\cite{PT-theory} and experimentally~\cite{PT-exper}, without any connection to the quantum theory, in the context of classical optics, as well as acoustics~\cite{acoustics}, microwaves~\cite{microw}, electronics~\cite{electronics}, and optomechanics~\cite{mechanical}, making use of the fundamental fact that the paraxial propagation equation, which is commonly used in optics, has essentially the same form as the quantum-mechanical Schr\"{o}dinger equation. Accordingly, the spatially even and odd Hermitian and anti-Hermitian terms of the underlying Hamiltonian correspond, respectively, to a symmetric spatial pattern of the local refractive index and antisymmetric distribution of local gain and loss in the waveguide.

Further, the presence of the Kerr nonlinearity, which is ubiquitous in optics, has suggested the consideration of Hamiltonians that include the corresponding quartic terms too. The nonlinearity readily gives rise to families of $\mathcal{PT}$-symmetric solitons, which have been explored in various contexts (see recent reviews~\cite{Reviews}). In particular, a natural setting for the prediction of such one- and two-dimensional solitons is provided by $\mathcal{PT}$-symmetric dual-core waveguides~\cite{PT-coupler}. Although the underlying setting is non-Hermitian, $\mathcal{PT}$-symmetric solitons exist in continuous families, like in conservative systems, rather than as isolated dissipative solitons.

The well-elaborated emulation of the non-Hermitian $\mathcal{PT}$ symmetry in optics suggests one to seek for a possibility to realize non-Hermitian Hamiltonians featuring other fundamental symmetries in appropriately designed optical settings, a natural candidate being the $\mathcal{CP}$ symmetry. This was proposed in Ref.~\cite{dana2015cp}, using a model of dual-core optical fibers, with opposite signs of the group-velocity dispersion (GVD) in the two cores and phase-velocity mismatch between them. The non-Hermitian ingredient of the system is the specific intercore coupling, which, in a phenomenological form, can represent gain and loss in the system, assuming that the coupler is embedded in an \textquotedblleft active" medium~\cite{bara13,Laksh}. Alternatively, the same coupling can be derived directly for two fundamental-frequency components of a nondegenerate (three-wave) second-harmonic-generating system, assuming that the depletion of the second-harmonic pump is negligible~\cite{dana2015cp}. In terms of this system, the $\mathcal{P}$ transform is realized as the swap of the two cores, and simultaneous inversion of the sign of the temporal variable in the transmission equations, while $\mathcal{C}$ amounts to the replacement of the wave amplitude by its complex-conjugate counterpart. The nonlinear version of the $\mathcal{CP}$-symmetric system, derived in Ref.~\cite{dana2015cp}, gives rise to a family of stable \textit{gap solitons}, even if the Kerr nonlinearity breaks the $\mathcal{CP}$ symmetry. A possibility to implement non-Hermitian $\mathcal{CP}$ symmetry in the context of matter waves was elaborated in terms of a two-component atomic Bose-Einstein condensate with the spin-orbit coupling between the components, assuming that one of them carries the gain and the other one is subject to the action of loss with the same strength~\cite{something}.

In this work, we aim to derive a discrete version of non-Hermitian $\mathcal{CP}$-invariant systems, which calls for implementation in terms of an appropriate optical system. The system is realized as an array of dual-core optical waveguides in the spatial domain, with the temporal-domain GVD replaced by the discrete diffraction~\cite{review} in two parallel guiding arrays of the system. While in dual-core fibers it is easy to realize the setting with opposite signs of the temporal GVD in parallel cores~\cite{Dave}, the implementation of opposite signs of the discrete diffraction is a challenging element of the model. As we discuss below, this can be realized by means of the diffraction-management technique~\cite{Silb}. We construct several species of fundamental discrete solitons in the framework of the obtained system, which includes the Kerr nonlinearity. Similar to the abovementioned $\mathcal{PT}$-invariant solitons, they exist here in continuous families, despite the non-Hermitian character of the system. The soliton families are obtained in an approximate analytical and full numerical forms, starting from the anticontinuum limit (uncoupled array). One family, constructed in the system's finite bandgap, continues, as a completely stable one, into the abovementioned gap solitons found in the continuum-limit variant of the system. Other families are found in semi-infinite gaps. They all terminate before reaching the continuum limit. One family features an internal boundary of oscillatory instability, all others being stable as long as they exist.

Previously, various species of one- and two-dimensional (1D and 2D) lattice solitons, such as 1D twisted modes~\cite{twisted} and discrete vortices~\cite{MK}, which may be (partly) stable in the discrete form, but vanish or suffer destabilization in the continuum limit, have been found in conservative models, such as the discrete nonlinear Schr\"{o}dinger equation (NLSE)~\cite{PGK}, but they have not been found in non-Hermitian systems. It is worthy to note that the all families revealed by the present analysis in semi-infinite gaps represent several species of fundamental solitons (on-site-centered single-peak ones), while the abovementioned twisted and vortex modes in conservative lattices are higher-order states. Further, it is relevant to stress too that, in the context of the discrete NLSE, the existence and stability of such discrete 1D and 2D states are controlled by the single effective parameter, viz., the relative strength of the intersite coupling, with respect to the strength of onsite nonlinearity~\cite{PGK}. On the other hand, the families of 1D discrete solitons, which are reported in the present work, may be better fitted to experimental settings, as their existence and stability are additionally controlled by the phase-velocity-mismatch and gain-loss parameters. In particular, the latter coefficient can be easily adjusted by varying the gain applied to the host active medium.

The manuscript is organized as follows. The model is introduced in Section~\ref{sec2model}. The perturbation theory, which makes use of weak couplings, is applied to fundamental discrete solitons in Section~\ref{sec3analytic}. In addition to the weak coupling between the sites (i.e., between $\mathcal{CP}$-symmetric dimers), the analysis is also performed for a small gain-loss coefficient, which accounts for the active coupling between the cores of the dimer elements. The existence and stability of the discrete solitons are then considered by means of numerical methods in Section~\ref{sec4}, finding stationary states and solving the eigenvalue problem for small perturbations around them. Results of the numerical calculations are compared to their analytical counterparts. In particular, we produce stability regions for the fundamental on-site solitons, which are controlled, as stated above, by both the intersite-coupling strength and the gain-loss parameter of the interdimer coupling, in addition to the intercore phase-velocity mismatch. We also explore dynamics of unstable solitons by means of direct simulations. The paper is concluded by Section~\ref{sec5}.

\section{The model} \label{sec2model}

The dimerized chain of couplers under the consideration is described by coupled equations for amplitudes $u_{n}$ and $v_{n}$ of electromagnetic waves in the coupled cores:
\begin{equation}
\begin{split}
\dot{u}_{n} &= i|u_{n}|^{2}u_{n} + i\epsilon \Delta_{2} u_{n} + \gamma v_{n} - iqu_{n}, \\
\dot{v}_{n} &= i|v_{n}|^{2}v_{n} - i\epsilon \Delta_{2} v_{n} + \gamma u_{n} + iqv_{n},
\end{split} \label{1}
\end{equation}
where the dot stands for the derivative with respect to evolution variable $z$, which is the propagation distance in the array of optical waveguides, the cubic terms represent the usual intracore Kerr nonlinearity, and $\epsilon > 0$ is the coefficient of the horizontal linear coupling with opposite signs, acting along each subchain between adjacent sites, $\Delta_{2} u_{n} = (u_{n + 1} - 2u_{n} + u_{n - 1})$ and $\Delta_{2} v_{n} = (v_{n + 1} - 2 v_{n} + v_{n - 1})$ being the respective finite-difference second derivatives, which represent the discrete spatial diffraction in the parallel arrays ($\epsilon < 0$ may be replaced by $\epsilon > 0$ simply by renaming $u_{n} \longleftrightarrow v_{n}$).

The opposite signs of the discrete diffraction in the two parallel arrays (with spacing $d$), which is an essential ingredient of the present system, may be realized by means of the \textit{diffraction-management} technique~\cite{Silb}, i.e., coupling into one of the arrays a light beam with a small perpendicular component $\kappa_{\perp} = \pi/(2d)$ of the wave vector, the corresponding discrete-diffraction coefficient being $\sim \cos \left(\kappa_{\perp} d \right)$. Another essential ingredient of the present system is the vertical coupling between the parallel arrays, represented by real coefficient $\gamma > 0$ ($\gamma < 0$ may be replaced by $\gamma > 0$, renaming $v_{n} \rightarrow -v_{n}$), which acts as the gain or loss in the active system~\cite{bara13}. The last terms in Eqs.~\eqref{1}, with coefficient $q \gtrless 0$, represent a phase-velocity mismatch between the cores. While $q$ may be scaled to be $\pm 1$, it is more convenient to keep it as a free parameter.

It is straightforward to check that the linearized version of Eqs.~\eqref{1} is symmetric under the abovementioned $\mathcal{CP}$ transformation $u_{n} \rightarrow v_{n}^{\ast}, \, v_{n} \rightarrow u_{n}^{\ast}$, where $^{\ast}$ stands for the complex conjugation; i.e., the linear system supports the ${\mathcal{CP}}$ symmetry, while the Kerr terms are not compatible with the transformation~\cite{dana2015cp}. Our objective is to construct discrete solitons of the full nonlinear system, subject to the localization conditions, $u_{n}, v_{n}\rightarrow 0$ as $n\rightarrow \pm \infty$.

The continuum limit of system~\eqref{1}, which corresponds to $\epsilon \rightarrow \infty $ and the discrete coordinate $n$ replaced by a continuous one, $x$, produces a system of coupled NLSEs: 
\begin{equation}
\begin{split}
\frac{\partial u}{\partial z} = i|u|^{2} u + i \frac{\partial^{2} u}{\partial x^{2}} + \gamma v - iqu, \\
\frac{\partial v}{\partial z} = i|v|^{2} v - i \frac{\partial^{2} v}{\partial x^{2}} + \gamma u + iqv.
\end{split} \label{cont}
\end{equation}
This system was investigated by means of analytical and numerical methods in Ref.~\cite{dana2015cp}. In the opposite (anticontinuum) limit, with $\epsilon = 0$~\cite{anti,Aubry}, the chain~\eqref{1} amounts to a set of isolated dimers with a complex intercore coupling. Such dimers with 2~degrees of freedom have been studied in detail in Ref.~\cite{bara13}.

Stationary solutions to Eqs.~\eqref{1} with real propagation constant $-K$ are sought for in the usual forms, 
\begin{equation}
u_{n} = A_{n} e^{-iKz}, \qquad v_{n} = B_{n} e^{-iKz},  \label{3.1a}
\end{equation}
with complex amplitudes $A_n$ and $B_n$ obeying the coupled algebraic equations:
\begin{equation}
\begin{split}
K A_{n} = -A_{n}^{2} A_{n}^{\ast} - \epsilon (A_{n + 1} - 2 A_{n} + A_{n - 1}) + i \gamma B_{n} + q A_{n}, \\
K B_{n} = -B_{n}^{2} B_{n}^{\ast} + \epsilon (B_{n + 1} - 2 B_{n} + B_{n - 1}) + i \gamma A_{n} - q B_{n}.
\end{split} \label{2}
\end{equation}
Using the invariance of Eqs.~\eqref{2} with respect to the phase shift, one can infer that localized stationary solutions can be found with real-valued $A_{n}$ and purely imaginary $B_{n}$. On the other hand, looking for solutions to the linearized version of Eqs.~\eqref{2} plane waves, $\left(A_{n}, B_{n} \right) = \left(A_{0}, B_{0} \right) \exp \left(ikn \right)$ with the real wave number $k$, we obtain the dispersion relation for the linearized system:
\begin{equation}
K^{2} = \left[q + 4 \epsilon \sin^{2} (k/2) \right]^{2} - \gamma^{2}. 		\label{disp}
\end{equation}
An essential corollary of Eq.~\eqref{disp} is that the stability of the zero solution, which plays the role of the background for bright solitons, holds under the condition $K^{2} \geq 0$, i.e.,
\begin{equation}
q \geq \gamma,  \label{q>}
\end{equation}
for positive $q$, and
\begin{equation}
q \leq -\left(4 \epsilon + \gamma \right),  \label{q<0}
\end{equation}
for negative $q$. These conditions demonstrate that the presence of the phase-velocity mismatch, $q \neq 0$, is necessary for the stability of localized states (recall that we have set $\gamma > 0$ and $\epsilon > 0$). The increase of the gain-loss coefficient, $\gamma$, from small values leads to the \emph{breaking} of the $\mathcal{CP}$ symmetry in the linearized system at critical points, $\gamma_{\mathrm{cr}} = q$ for $q > 0$, and at $\gamma_{\mathrm{cr}} = |q| - 4 \epsilon $ (provided that $|q|$ exceeds $4 \epsilon$ in the latter case, otherwise the $\mathcal{CP}$ symmetry is always broken).

If condition~\eqref{q>} holds, the existence of discrete solitons may be expected in spectral \textit{band gaps}, i.e., intervals of values of $K^2$ which cannot be covered by Eq.~\eqref{disp} with $\sin^{2} (k/2) \leq 1$. These are finite and semi-infinite band gaps, viz.,
\begin{equation}
K^2 < q^2 - \gamma^2 \qquad \text{or} \qquad K^2 > \left(q + 4 \epsilon \right)^2 - \gamma^2,  \label{ww}
\end{equation}
in the case defined by Eq. (\ref{q>}), and
\begin{equation}
K^2 < \left(q + 4 \epsilon \right)^2 - \gamma^2 \qquad \text{or} \qquad K^2 > q^2 - \gamma^2,  \label{ww2}
\end{equation}
in the case of Eq.~\eqref{q<0}. Note that, in the continuum limit, which is represented by Eq.~\eqref{cont}, the stability condition for the zero solution is given by Eq.~\eqref{q>} [while Eq.~\eqref{q<0} is obviously irrelevant in this limit], and the respective band gap is the finite one, defined by the first inequality in Eq.~\eqref{ww}~\cite{dana2015cp}, while the semi-infinite band gap is expelled to infinitely large values of $K^2$.

It is relevant to stress that the identification of the band gaps as the habitat for solitons in the non-Hermitian system is not self-obvious. Nevertheless, this principle, suggested by studies of conservative systems, is valid, as long as the spectrum remains completely real, i.e., the $\mathcal{CP}$ symmetry is not broken, being secured by Eqs.~\eqref{q>} and~\eqref{q<0}. The same is true for solitons in $\mathcal{PT}$-symmetric systems~\cite{Reviews}.

To investigate stability of stationary states against perturbations with an infinitesimal real amplitude $\zeta$, the perturbed solution is defined as 
$u_{n} = [A_{n} + \zeta (Q_{n} + i R_{n}) e^{\lambda z}] e^{-iKz}$ and $v_{n} = [B_{n} + \zeta (S_{n} + i T_{n}) e^{\lambda z}] e^{-iKz}$, where eigenvalue $\lambda$ should be found from a numerical solution of the system of linearized equations for real form-factors $Q_{n}$, $R_{n}$ and $S_{n}$, $T_{n}$, in which it is taken into regard that amplitudes $A_{n}$ and $B_{n}$ are real and purely imaginary, respectively, as stated above:
\begin{equation}
\begin{split}
\lambda Q_n &= -(  A_n^2 + K - q) R_n - \epsilon (R_{n + 1} - 2 R_n + R_{n - 1}) + \gamma S_n, \\
\lambda R_n &=  (3 A_n^2 + K - q) Q_n + \epsilon (Q_{n + 1} - 2 Q_n + Q_{n - 1}) + \gamma T_n, \\
\lambda S_n &=  (3 B_n^2 - K - q) T_n + \epsilon (T_{n + 1} - 2 T_n + T_{n - 1}) + \gamma Q_n, \\
\lambda T_n & = (- B_n^2 + K + q) S_n - \epsilon (S_{n + 1} - 2 S_n + S_{n - 1}) + \gamma R_n.
\end{split}			\label{8a}
\end{equation}
As usual, the stationary solution is linearly stable if the condition $\mathrm{Re}(\lambda) \leq 0$ holds for all eigenvalues, and is unstable otherwise.

\section{Analytical calculations} 		\label{sec3analytic}

\subsection{The anticontinuum limit} 	\label{dimers}

In the decoupled array, with $\epsilon = 0$, stationary solutions of Eqs.~\eqref{2} can be written as $A_n^{(0)} = \tilde{a}_0$ and $B_n^{(0)} = i \tilde{b}_0$, with real $\tilde{a}_0$ and $\tilde{b}_0$. Upon substitution of this into Eqs.~\eqref{2}, one obtains
\begin{equation}
\tilde{b}_0 = \left(\tilde{a}_0/\gamma \right) \left[-\tilde{a}_0^2 + (q - K) \right] ,  \label{b0}
\end{equation}
where $\tilde{a}_{0}$ solves the polynomial equation
\begin{align}
\tilde{a}_0^9 - 3(q - K) \tilde{a}_0^7 + 3 (q - K)^2 \tilde{a}_0^5 \nonumber \\ 
+ [\gamma^2 (q + K) - (q - K)^3] \tilde{a}_0^3 \nonumber \\ + [\gamma^4 - \gamma^2 (q^2 - K^2)] \tilde{a}_0 = 0.					\label{pp}
\end{align}

One solution of Eq.~\eqref{pp} is a trivial one, $\tilde{a}_0 = \tilde{b}_0 = 0$, nontrivial solutions for $\tilde{a}_0^2$ being roots of a quartic polynomial, which can be formally solved in an analytical form, producing, however, impractically cumbersome expressions~\cite{nick09}. The analysis of Eq.~\eqref{pp} simplifies for small values of the intercore coupling, $\gamma$, and $q$ close to $\pm 1$, viz.,
\begin{equation}
q = \pm 1 - \hat{q} \gamma,  \label{+-}
\end{equation}
with $\hat{q} \sim 1$. First, for $q = +1 - \hat{q} \gamma$, expanding Eq.~\eqref{pp} up to $\mathcal{O}(\gamma^2)$, we find four relevant roots:
\begin{align}
\tilde{a}_0 &= \frac{2(1 - K) - \gamma \hat{q}}{2\sqrt{1 - K}} + \dots, \quad  
\tilde{b}_0  = \frac{\sqrt{1 - K} \gamma}{1 + K} + \dots,  					\label{a0b0_3rd} \\
\tilde{a}_0 &= \frac{\sqrt{-\left(1 + K \right)} \gamma}{1 - K} + \dots, \quad 
\tilde{b}_0  = \sqrt{-\left(1 + K \right)} + \dots,  						\label{a0b0_4thc} \\
\tilde{a}_0 &= \sqrt{1 - K} \pm \frac{\sqrt{-(1 + K)} - \hat{q} \gamma \sqrt{1 - K}}{2(1 - K)} + \dots, \nonumber \\ 
\tilde{b}_0 &= \pm \sqrt{-(1 + K)} + \dots,	 \qquad  						\label{a0b0_12stnd}
\end{align}
which exist at $K < +1$, $K < -1,$ and $K < -1$, respectively. Similarly, for $q = -1 - \hat{q} \gamma$ we also find four roots:
\begin{align}
\tilde{a}_0 &= -\frac{2(K + 1) + \hat{q} \gamma}{2\sqrt{-\left(1 + K \right)}} + \dots, \qquad   
\tilde{b}_0  = 0,  															\label{a0b0_3rd2} \\
\tilde{a}_0 &= -\frac{\sqrt{1 - K} \gamma}{1 + K} + \dots, \qquad  
\tilde{b}_0  = \sqrt{1 - K} + \dots,  										\label{a0b0_4th2} \\
\tilde{a}_0 &= \sqrt{-(1 + K)} \pm \frac{\sqrt{1 - K} - \hat{q} \gamma \sqrt{-(1 + K)}}{2(1 - K)} + \dots, \nonumber \\ 
\tilde{b}_0 & = \mp \sqrt{1 - K} + \dots, \qquad  							\label{a0b0_12stnd2}
\end{align}
which exist at $K < -1$, $K < +1$, and $K < -1$, respectively.

\subsection{Discrete solitons in the weakly coupled arrays} 				\label{dswc}

Because solutions $\tilde{a}_0$ and $\tilde{b}_0$ at each site $n$ are mutually independent in the decoupled array, one can construct infinitely many combinations, using different solutions for $\tilde{a}_0$ and $\tilde{b}_0$. Here, we focus on fundamental bright solitons of the on-site-centered type in the case of weak coupling, i.e., small $\epsilon$, which can be constructed by the continuation of the modes available at $\epsilon = 0$. This is a well-known method for finding various modes in discrete systems, starting from the anticontinuum limit~\cite{PGK}. Up to order $\epsilon^2$, such solitons are constructed in an approximate form: 
\begin{equation}
\begin{split}
A_{n} &= \left\{
\begin{array}{lll}
\tilde{a}_0 + \epsilon \tilde{a}_{0,1}, \qquad & n = 0, &  \\
			  \epsilon \tilde{a}_{1,1},        & n = \pm 1, &  \\
									 0, 	   & n \neq 0, \pm 1, &
\end{array}
\right. \\ 
B_{n} &= \left\{
\begin{array}{lll}
i\tilde{b}_0 + i \epsilon \tilde{b}_{0,1}, \qquad & n = 0, &  \\
			   i \epsilon \tilde{b}_{1,1}, 		  & n = \pm 1, &  \\
										0, 		  & n \neq 0,\pm 1, &
\end{array}
\right. 
\label{ds}
\end{split}  											
\end{equation}
where $\tilde{a}_0, \, \tilde{b}_0 \neq 0$ is one of the nonzero pairs given by Eqs.~\eqref{a0b0_3rd}--\eqref{a0b0_12stnd2}, and the next-order terms are obtained perturbatively from Eqs.~\eqref{2}, following the lines of Ref.~\cite{kiri16}:
\begin{equation}
\begin{split}
\tilde{a}_{0,1} = \frac{2 \gamma \tilde{b}_0 + 2 \tilde{a}_0 (q + K + 3\tilde{b}_0^2)}{\gamma^2 - (q - K - 3 \tilde{a}_0^2)(q + K + 3 \tilde{b}_0^2)}, \\
\tilde{b}_{0,1} = \frac{2 \gamma \tilde{a}_0 + 2 \tilde{b}_0 (q - K - 3\tilde{a}_0^2)}{\gamma^2 - (q - K - 3 \tilde{a}_0^2)(q + K + 3 \tilde{b}_0^2)}, \\
\end{split}
\end{equation}
\vspace*{-0.6cm}
\begin{equation}
\tilde{a}_{1,1} = \frac{  \gamma \tilde{b}_0 -   \tilde{a}_0 (q + K)}{\gamma^2 - (q^2 - K^2)}, 
\tilde{b}_{1,1} = \frac{ -\gamma \tilde{a}_0 -   \tilde{b}_0 (q - K)}{\gamma^2 - (q^2 - K^2)}. 
\end{equation}
\begin{figure*}[tbph]
\centering
\subfigure[]{\includegraphics[width=0.35\textwidth]{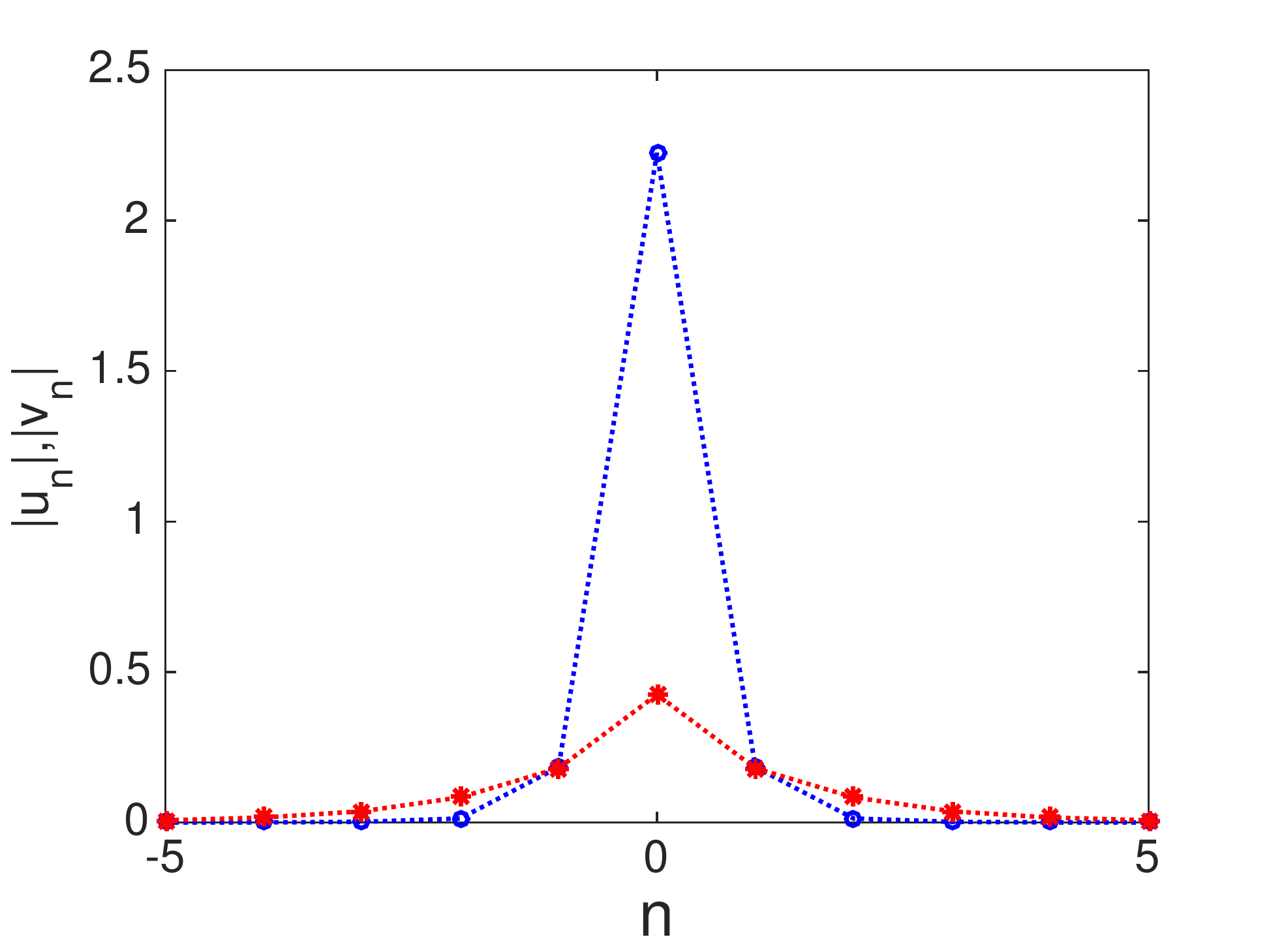}} 	\hspace{1cm}
\subfigure[]{\includegraphics[width=0.35\textwidth]{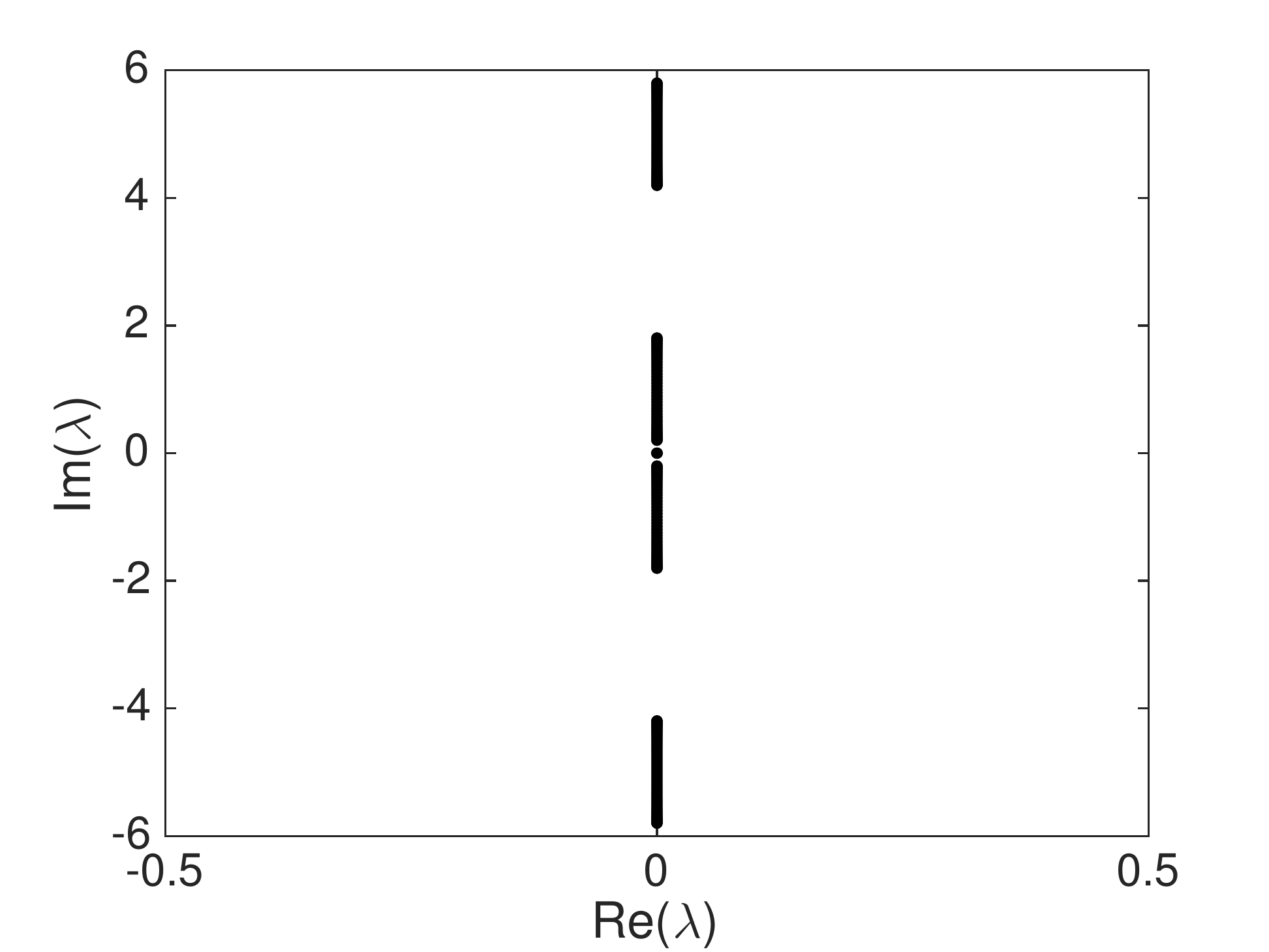}} 	\\
\subfigure[]{\includegraphics[width=0.35\textwidth]{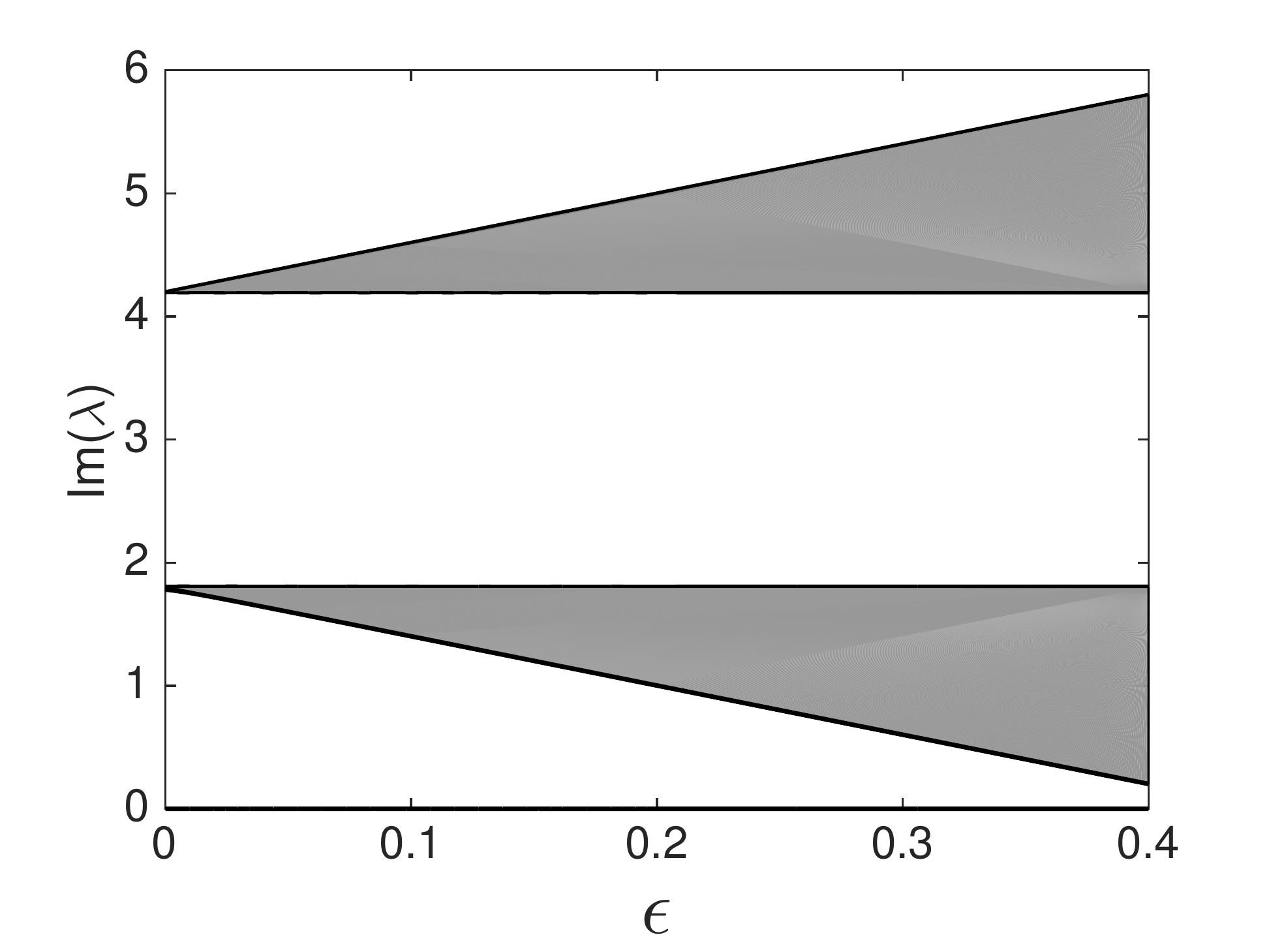}} \hspace{1cm}
\subfigure[]{\includegraphics[width=0.35\textwidth]{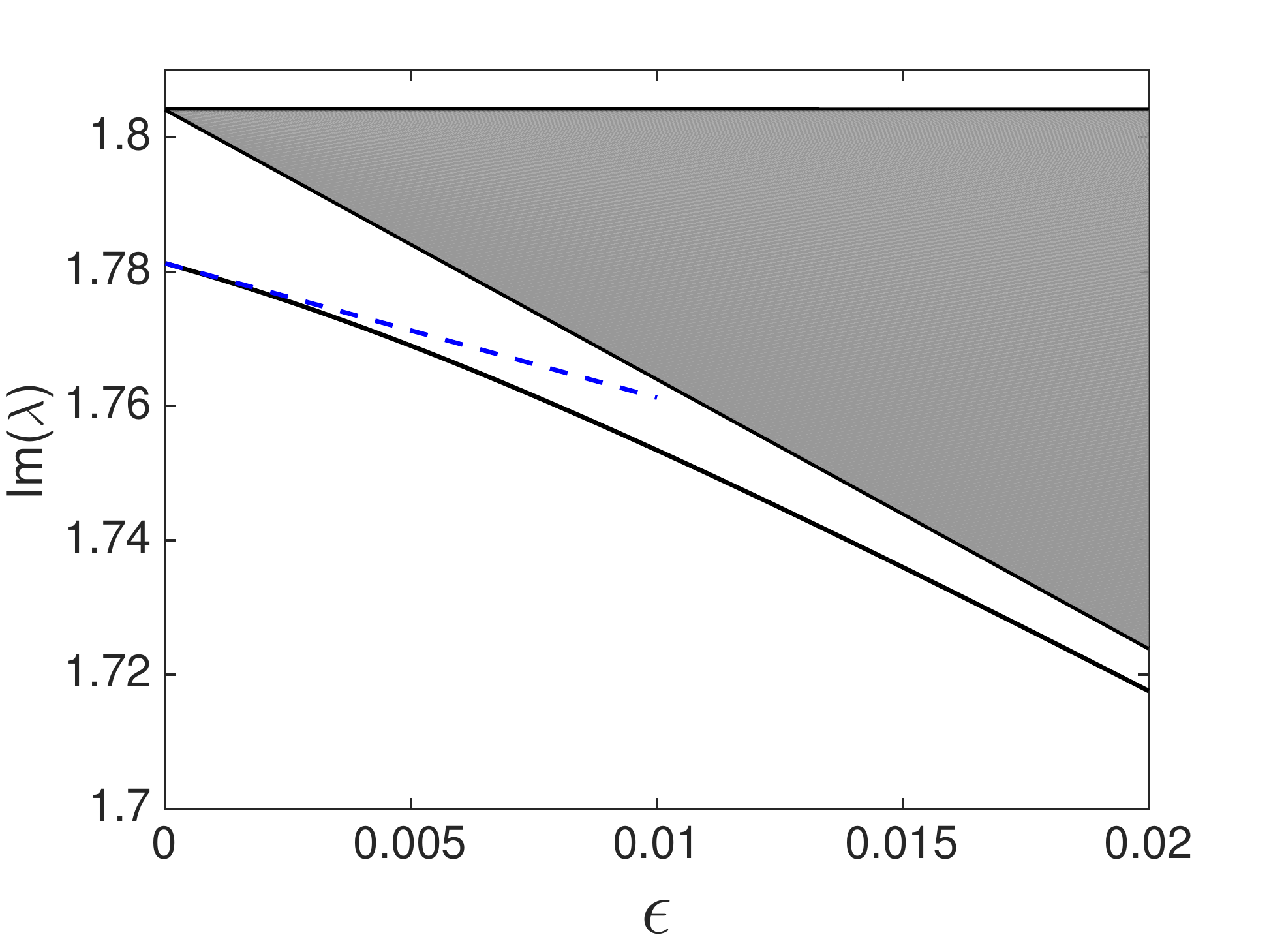}}
\caption{The stable discrete-soliton family initiated, in the analytical approximation, by Eqs.~\eqref{ds} and~\eqref{a0b0_3rd}, and its stability for $K = -3$, $\gamma = 0.1$, and $q = 1.2$. (a) The solution profile for $\epsilon = 0.4$ with the taller (blue) and shorter (red) curves corresponding to $|u_n|$ and $|v_n|$, respectively. (b) The corresponding spectrum of stability eigenvalues in the complex plane. (c) Imaginary eigenvalues (i.e., stable ones) as a function of $\epsilon $ [one branch is shown, the other one being its mirror image, cf. panel (b)]. (d) Zoom-in of panel (c) showing the separate eigenvalue initiated in the anticontinuum limit by the analytical approximation based on Eq.~\eqref{app1} (the approximation is displayed by the dashed line).} \label{3rdcase1} 
\end{figure*}

\vspace*{-0.5cm}
\subsection{Stability eigenvalues of the discrete solitons}				\label{sds}

In the framework of the weak-coupling limit elaborated in Sections~\ref{dimers} and~\ref{dswc}, we implement similar asymptotic expansions to solve semianalytically the stability-eigenvalue problem based on Eqs.~\eqref{8a}, i.e., we substitute in those equations
\begin{equation}
X = X^{(0)} + \sqrt{\epsilon} X^{(1)} + \epsilon X^{(2)} + \dots,  \label{box}
\end{equation}
with $X \equiv \left\{\lambda, Q_{n}, R_{n}, S_{n}, T_{n}\right\}$. Assuming the presence of the second independent small parameter, $\gamma$ (the intercore coupling parameter), coefficients in Eq.~\eqref{box} are further expanded as
\begin{equation}
X^{(j)} = X^{(j,0)} + \gamma X^{(j,1)} + \gamma^2 X^{(j,2)} + \dots,
\end{equation}
$j = 0, 1, 2, \dots$.~Details of the respective calculations are not shown here, as they directly follow the method elaborated in Ref.~\cite{kiri16}. Below, we report final results produced by this approach. It is relevant to stress that, while the expansion in terms of the small intersite coupling constant is a well-known approach, which has been elaborated for many conservative systems~\cite{anti,Aubry,PGK}, the analysis for non-Hermitian systems is developed here, and the use of the expansion in terms of two small parameters is an essential technical peculiarity, which may occur in the analysis of other non-Hermitian systems.

Due to the phase invariance, perturbation modes around the discrete solitons have a trivial eigenvalue $\lambda = 0$. In the case of $q = +1 - \hat{q}\gamma$ [see Eq.~\eqref{+-}], the discrete soliton~\eqref{ds}, with $\tilde{a}_0$ and $\tilde{b}_0$ taken as per Eqs.~\eqref{a0b0_3rd}, has a nonzero eigenvalue given, in the present approximation, by
\begin{equation}
\vspace*{-0.8pt}
\lambda = i \left[(1 + K) - \hat{q} \gamma + \mathcal{O}(\gamma^2) \right] + i \epsilon \left[2 + \mathcal{O}(\gamma^2) \right] + \mathcal{O}(\epsilon^{3/2}),  \label{app1}
\end{equation}
while for $\tilde{a}_0$ and $\tilde{b}_0$ taken as per Eqs.~\eqref{a0b0_4thc}, a nonzero stability eigenvalue is
\begin{equation}
\vspace*{-0.8pt}
\lambda = i \left[(-1 + K) + \hat{q} \gamma + \mathcal{O}(\gamma^2) \right] -i \epsilon \left[2 + \mathcal{O}(\gamma^2) \right] + \mathcal{O}(\epsilon^{3/2}).  \label{app2}
\end{equation}
In the case of $q = -1 - \hat{q} \gamma$, the discrete soliton~\eqref{ds}, with $\tilde{a}_0$ and $\tilde{b}_0$ taken as per Eqs.~\eqref{a0b0_3rd2}, has a nonzero eigenvalue given by
\begin{equation}
\vspace*{-0.8pt}
\lambda = i \left[(-1 + K) - \hat{q} \gamma + \mathcal{O}(\gamma^2) \right] + i \epsilon \left[2 + \mathcal{O}(\gamma^2) \right] + \mathcal{O}(\epsilon^{3/2}),  \label{app11}
\end{equation}
while, for $\tilde{a}_0$ and $\tilde{b}_0$ taken as per Eqs.~\eqref{a0b0_4th2}, it is
\begin{equation}
\vspace*{-0.8pt}
\lambda = i \left[(1 + K) + \hat{q} \gamma + \mathcal{O}(\gamma^2) \right] -i \epsilon \left[2 + \mathcal{O}(\gamma^2) \right] + \mathcal{O}(\epsilon^{3/2}).  \label{app22}
\end{equation}
In the present approximation, we conclude that the discrete solitons are stable, as all the corresponding eigenvalues are imaginary.

In the same approximation, it is not possible to produce nontrivial eigenvalues for the discrete soliton with $\tilde{a}_0$ and $\tilde{b}_0$ given by Eqs.~\eqref{a0b0_12stnd} and~\eqref{a0b0_12stnd2}, because, in both cases defined by Eq.~\eqref{+-} with small $\gamma$, the situation turns out to be degenerate, with all the eigenvalues remaining equal to zero.

\vspace*{-0.5cm}
\section{Numerical results}					\label{sec4}

Proceeding to the numerical analysis, we solved steady-state equations, Eqs.~\eqref{2}, by means of the Newton-Raphson method and then explored the stability of the numerical solutions by solving the eigenvalue problem~\eqref{8a}. Below, we present the numerical results as well as their comparison with the analytical calculations presented above.
\begin{figure*}[tbph]
\centering
\includegraphics[width=0.35\textwidth]{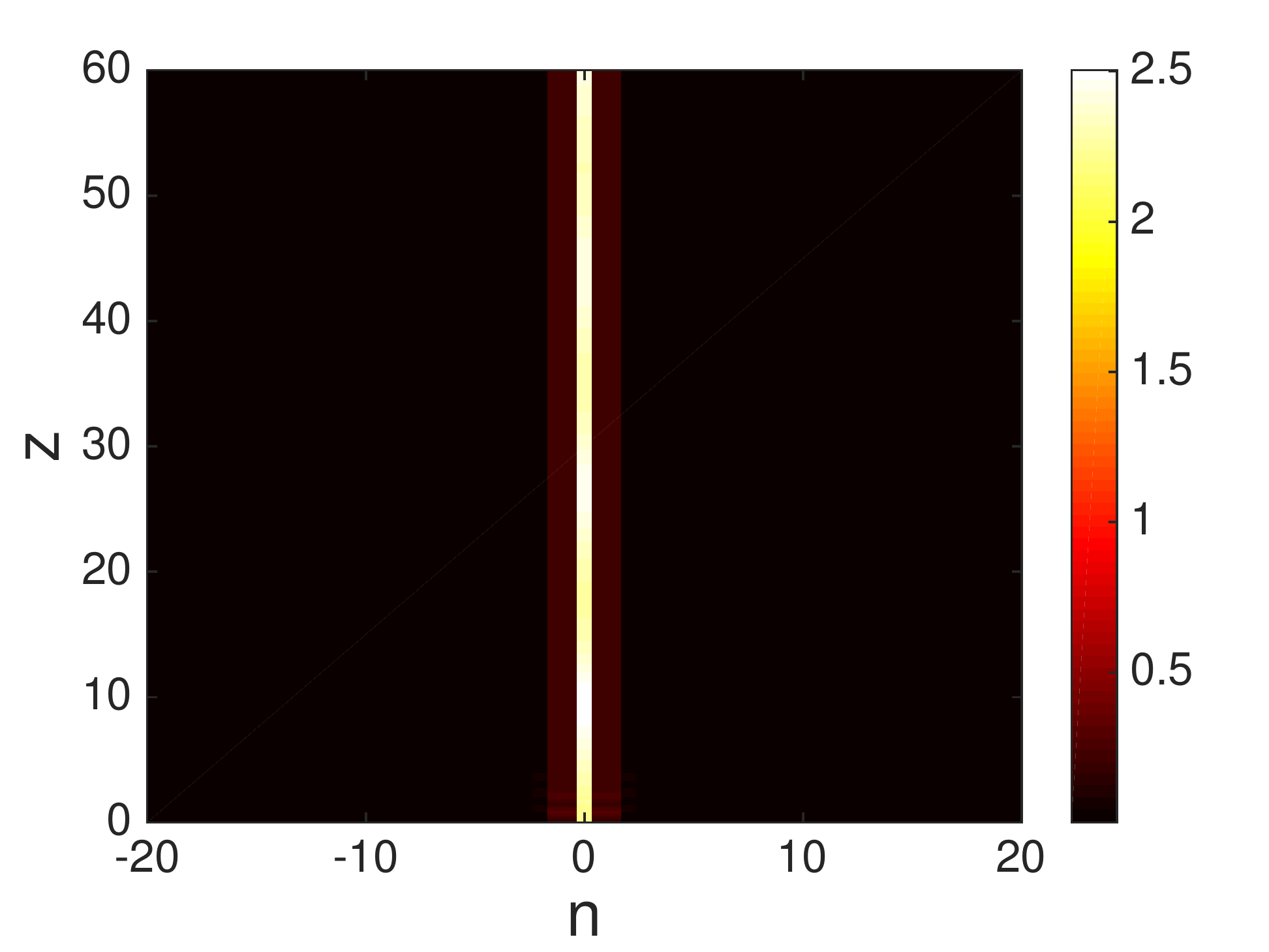} \hspace{1cm}%
\includegraphics[width=0.35\textwidth]{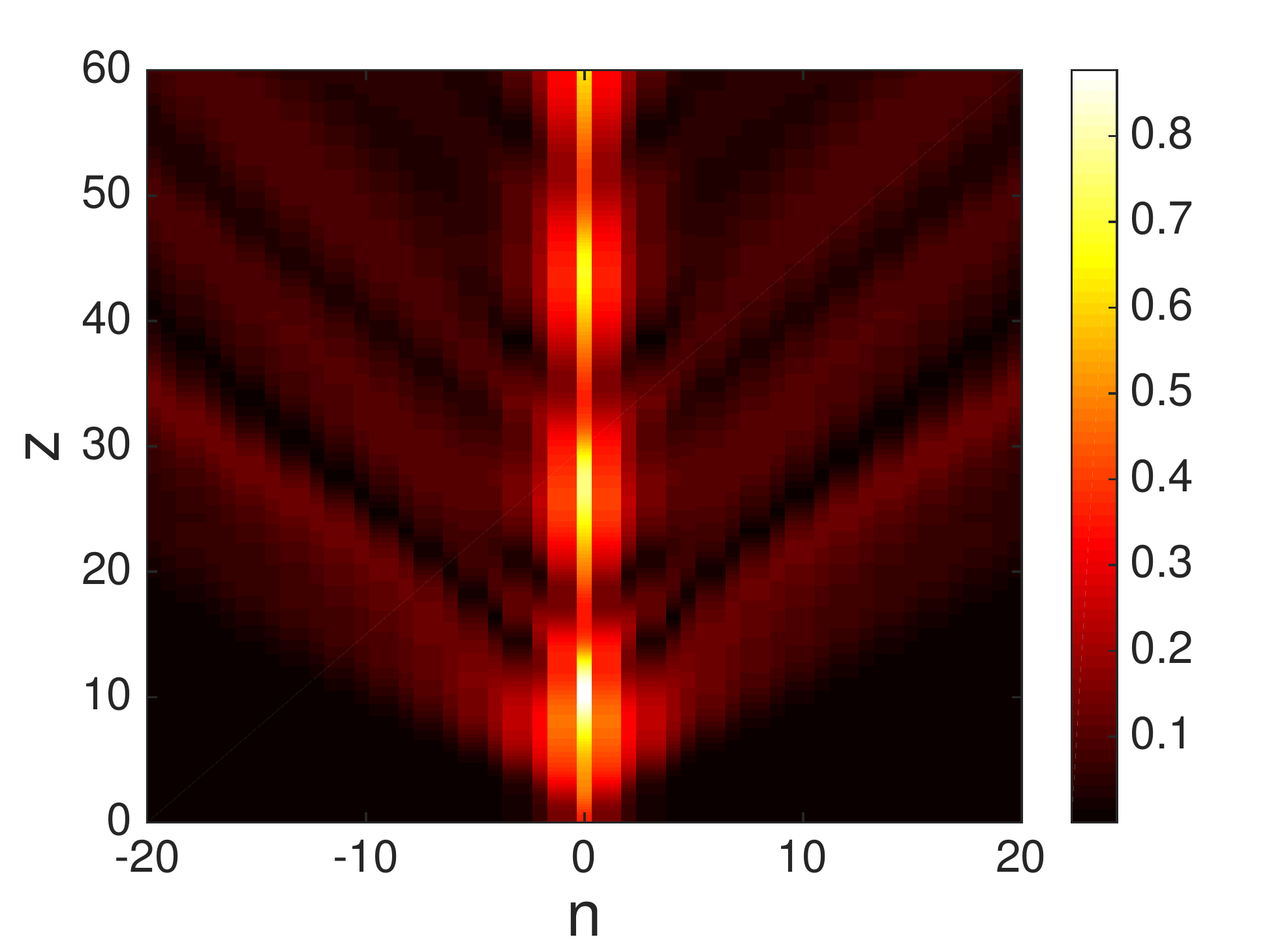} 
\caption{The evolution of a discrete soliton, corresponding to the configuration displayed in Fig.~\ref{3rdcase1}(a) beyond the critical value of the coupling constant, viz., at $\epsilon = 0.5$. Depicted in the left and right panels is the evolution of discrete fields $|u_{n}|^{2}$ and $|v_{n}|^{2}$, respectively.}
\label{Dynamic_case_3}
\end{figure*}
\begin{figure*}[tbph]
\centering
\subfigure[]{\includegraphics[width=0.35\textwidth]{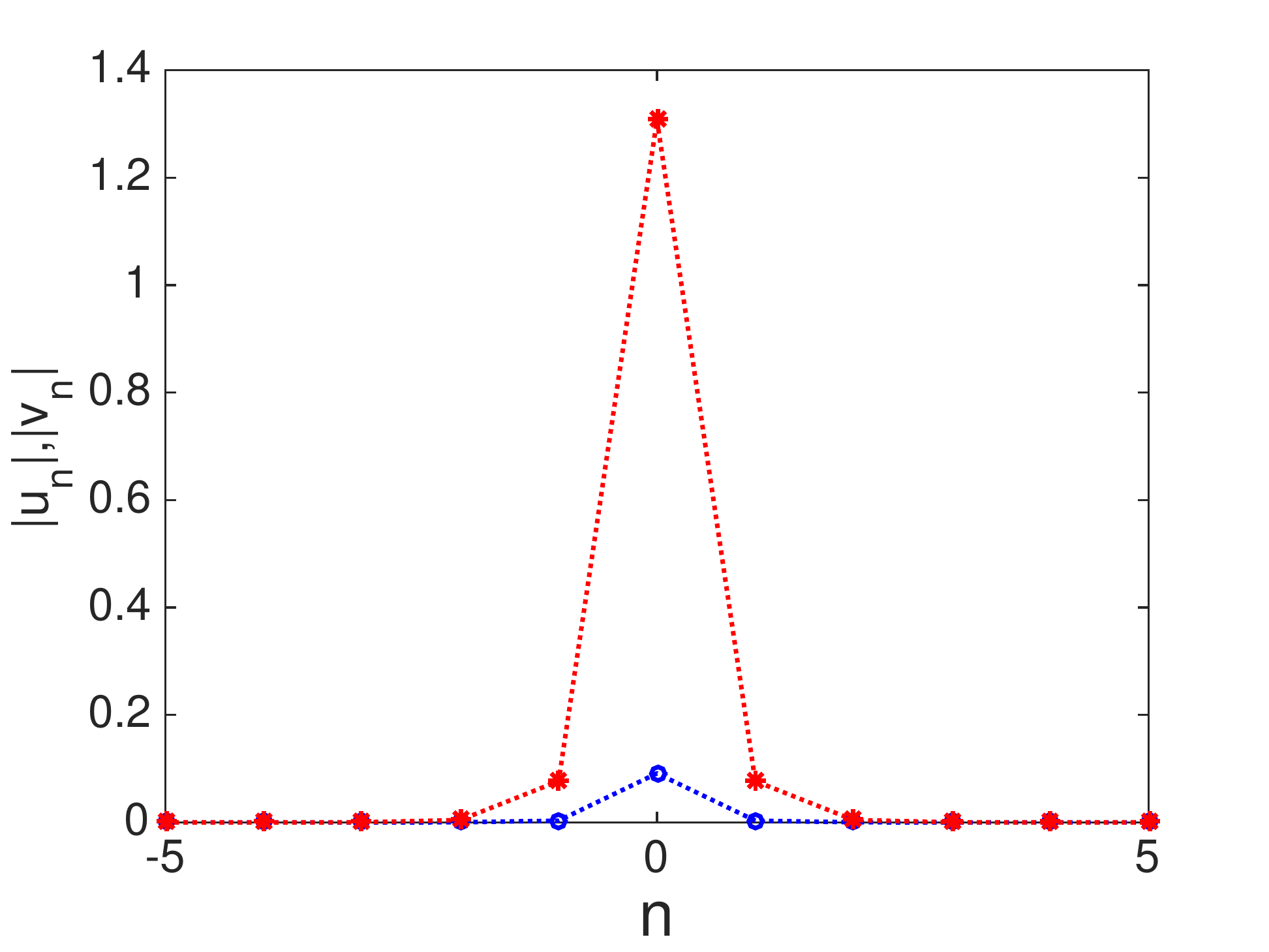}}  \hspace{1cm}
\subfigure[]{\includegraphics[width=0.35\textwidth]{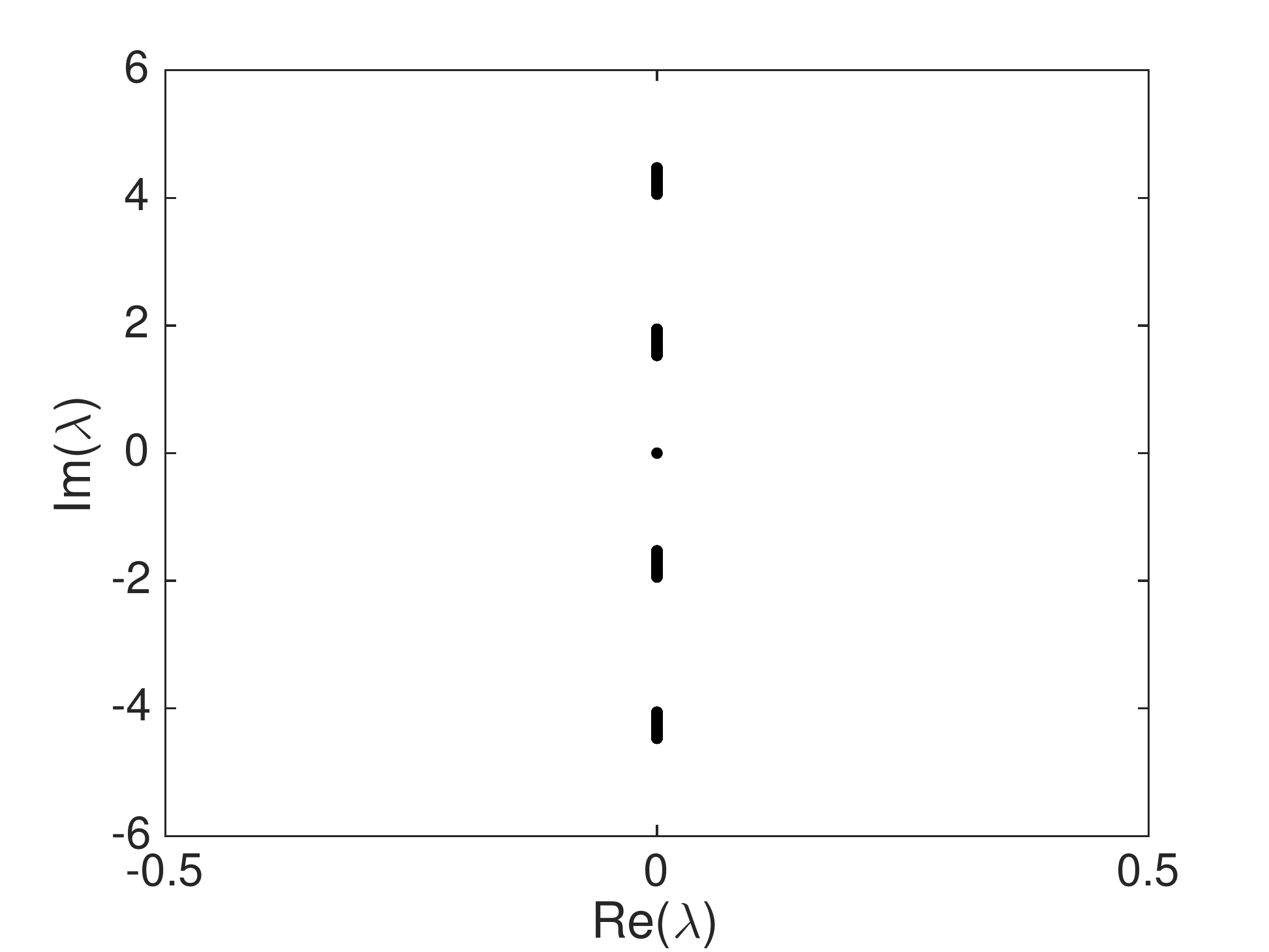}}  \\
\subfigure[]{\includegraphics[width=0.35\textwidth]{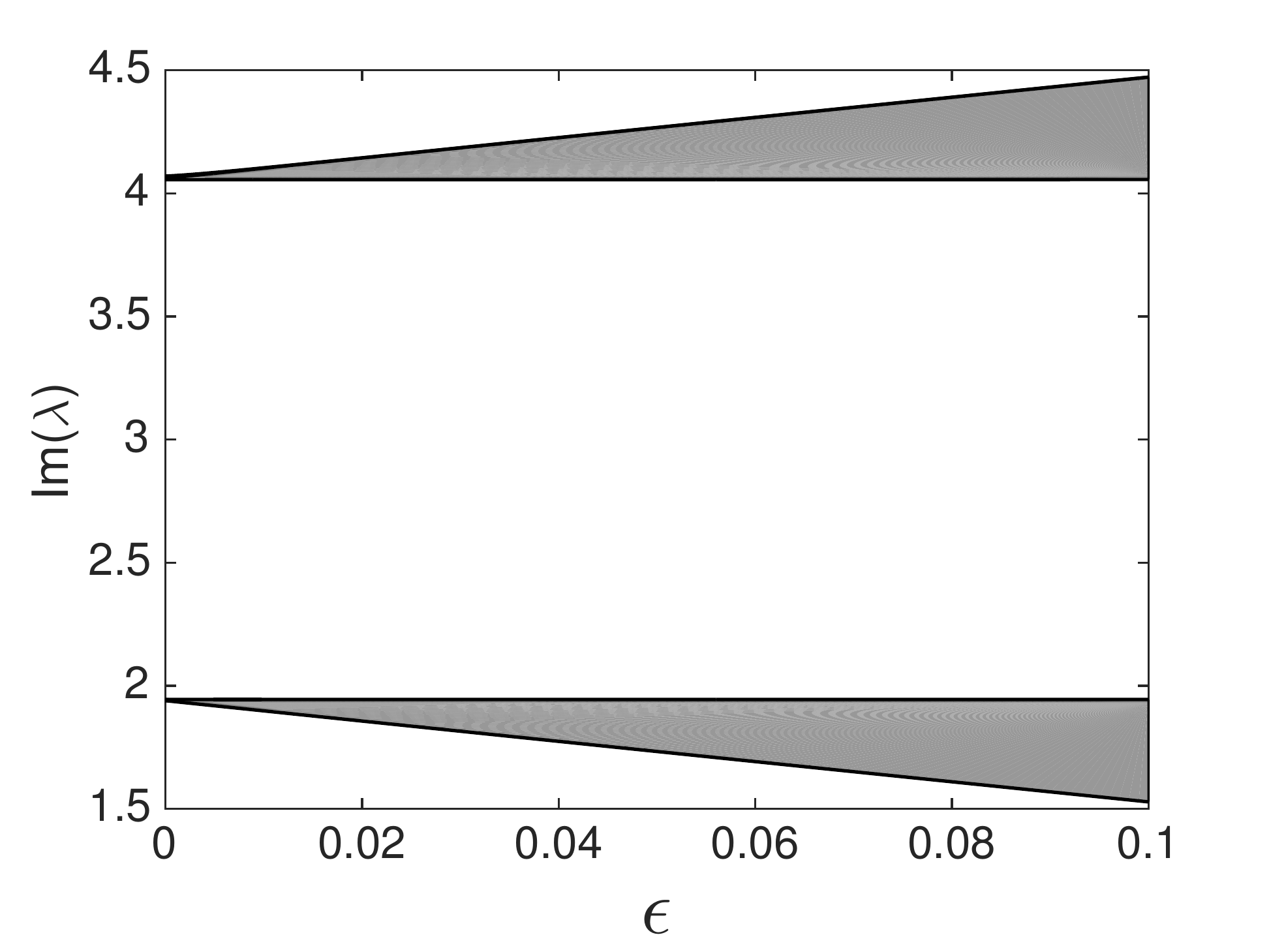}} \hspace{1cm}
\subfigure[]{\includegraphics[width=0.35\textwidth]{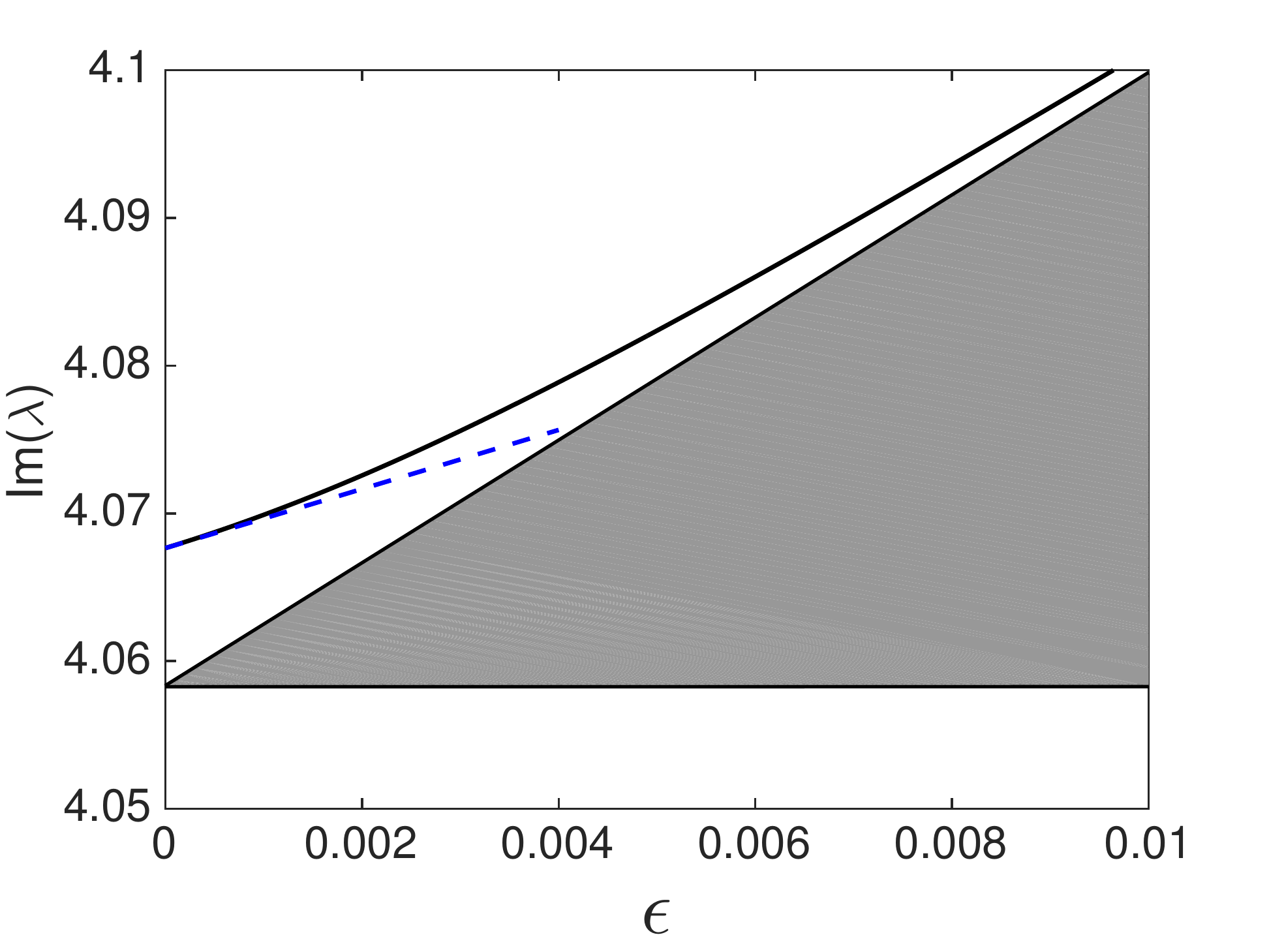}}
\caption{The same as Fig.~\ref{3rdcase1}, but for the fundamental discrete soliton given in the approximate analytical form by Eqs.~\eqref{ds} and~\eqref{a0b0_4thc}, and the set of its stability eigenvalues, for $K = -3$, $\gamma = 0.3$, and $q = 1.1$. In panel (a), the shorter (blue) and taller (red) curves correspond to $|u_n|$ and $|v_n|$, respectively. The dashed line in panel (d) represents the separate eigenvalue, as given by the analytical approximation~\eqref{app2}.}
\label{4thcase1}
\end{figure*}
\begin{figure*}[tbph]
\centering
\subfigure[]{\includegraphics[width=0.32\textwidth]{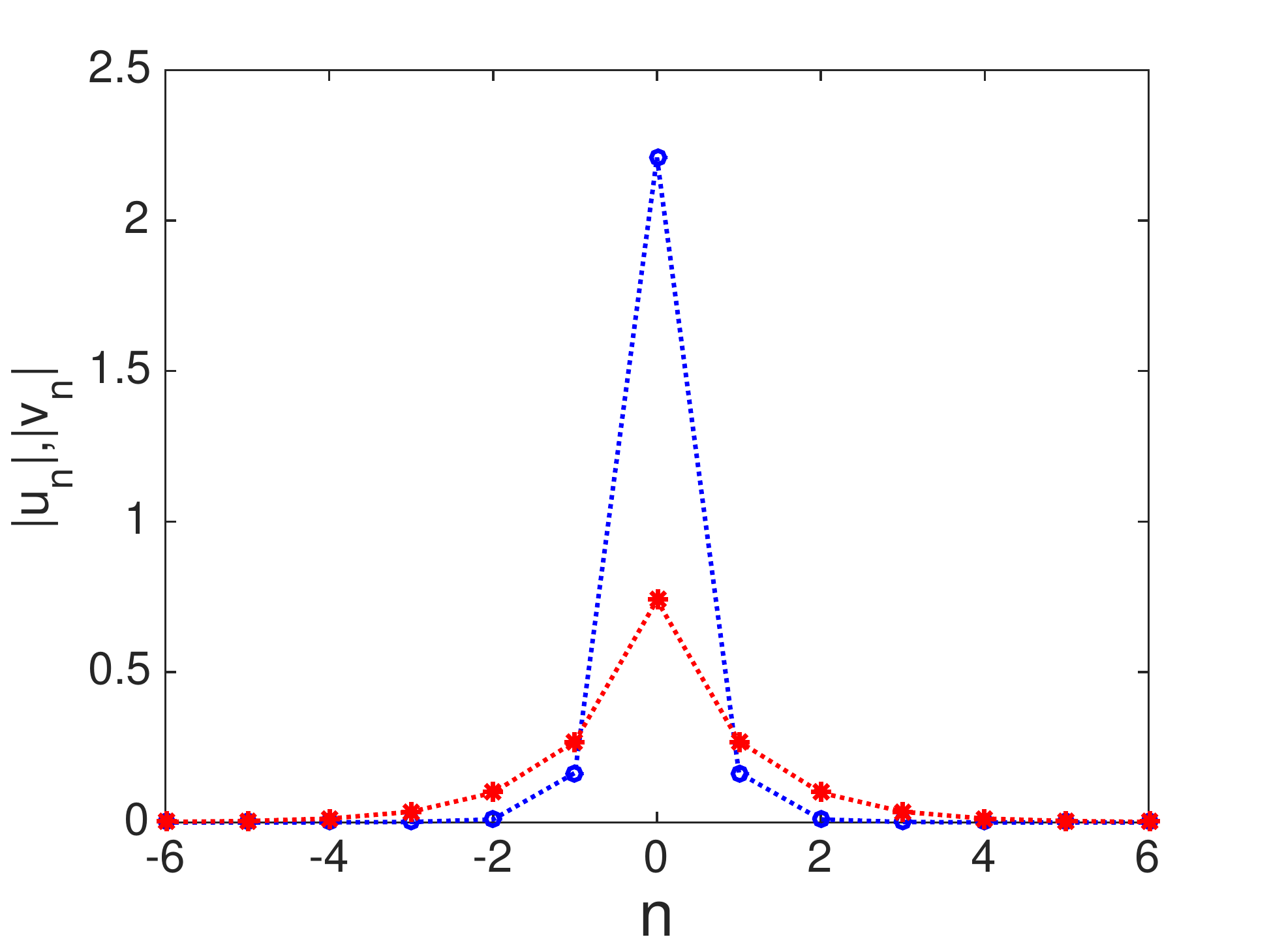}}
\subfigure[]{\includegraphics[width=0.32\textwidth]{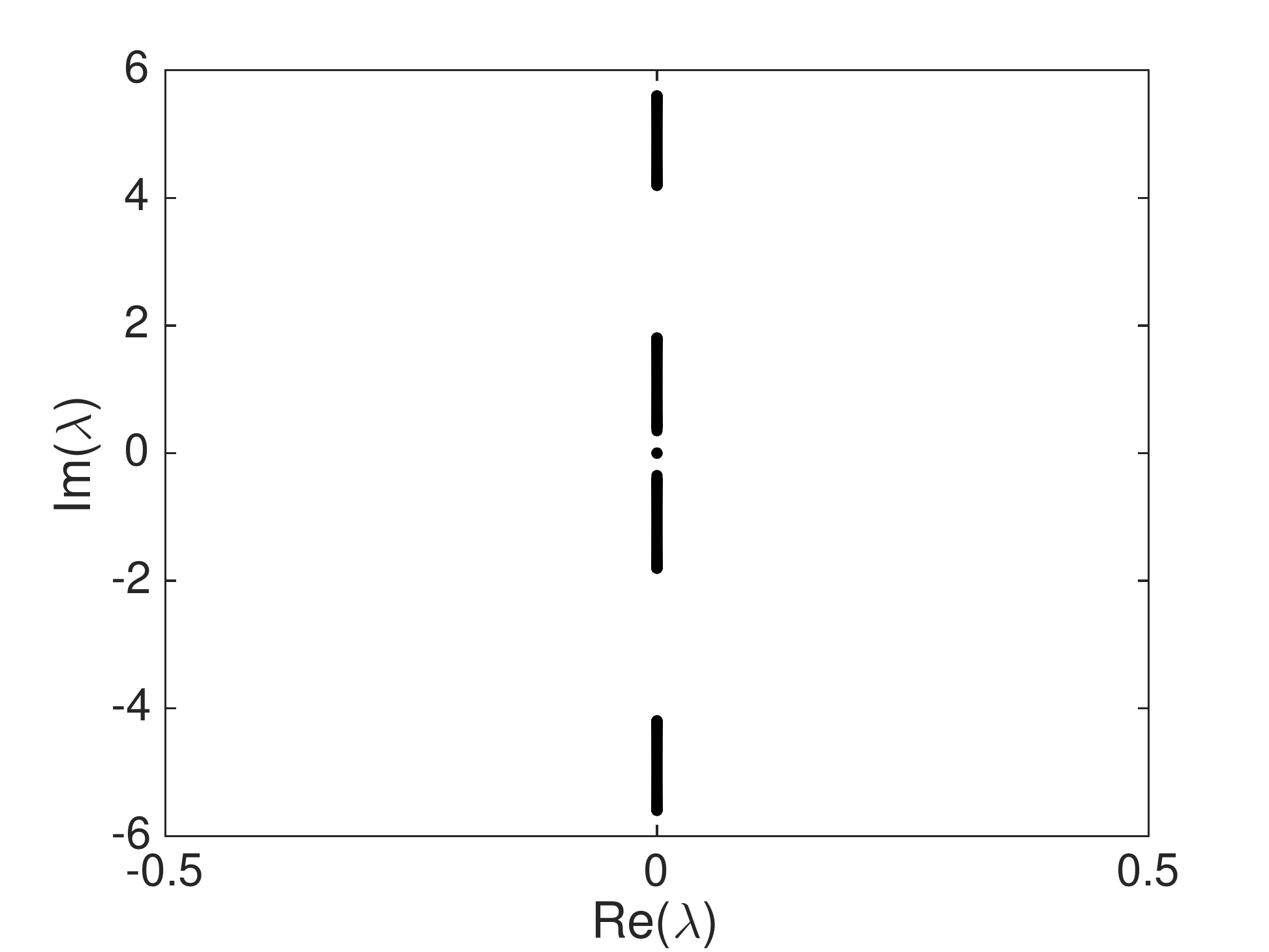}} 
\subfigure[]{\includegraphics[width=0.32\textwidth]{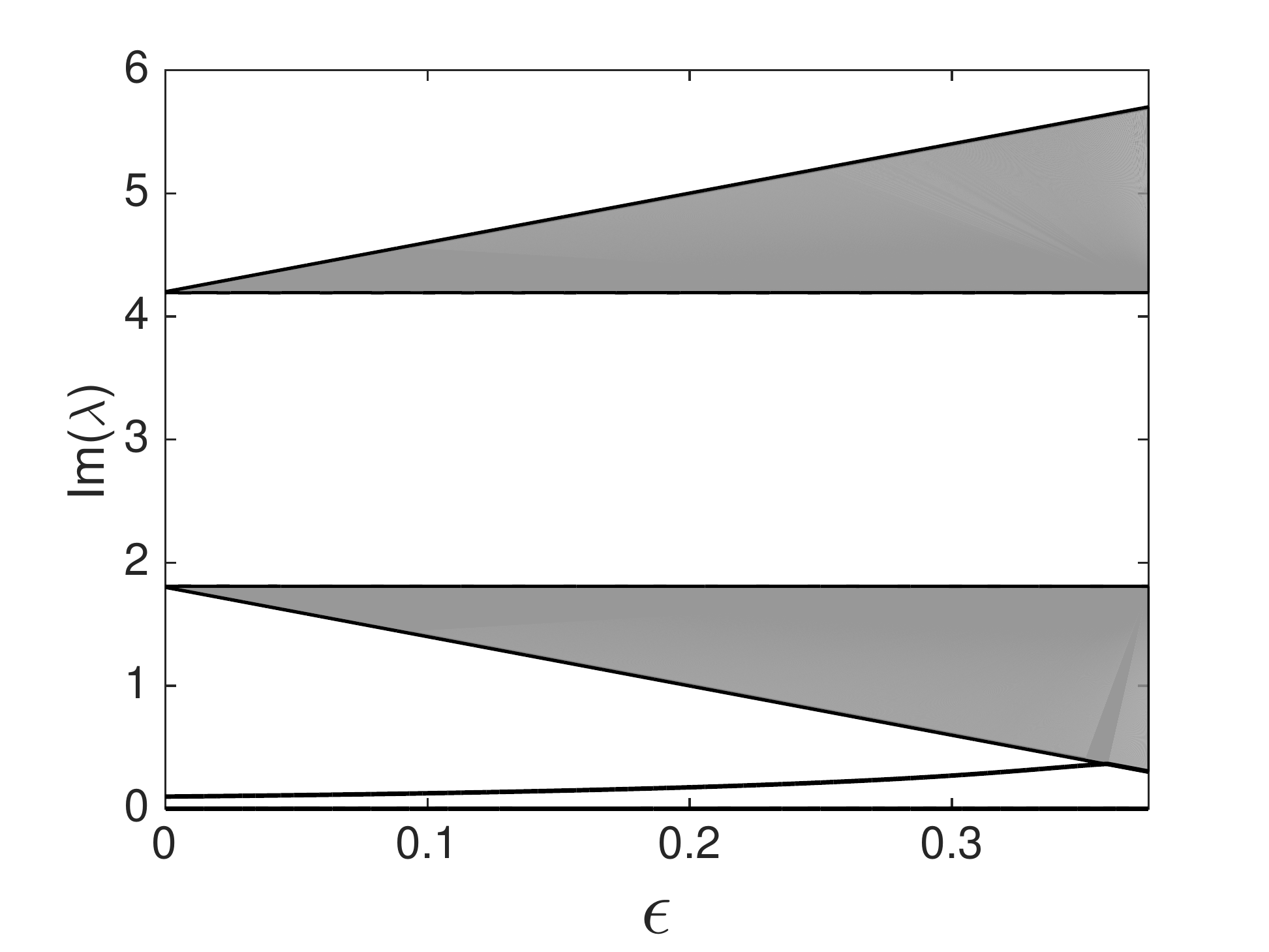}}
\caption{Discrete \textit{out-of-phase} fundamental solitons initiated by the analytical approximation based on Eqs.~\eqref{ds} and~\eqref{a0b0_12stnd}, with the ``$-$'' sign in the expressions of $\tilde{a}_0$ and $\tilde{b}_0$. Parameters are $K = -3$, $\gamma = 0.1$, and $q = 1.2$. (a)~The solution profile for~$\epsilon = 0.35$ with the taller (blue) and shorter (red) curves corresponding to $|u_n|$ and $|v_n|$, respectively. (b)~The spectrum of the corresponding stability eigenvalues in the complex plane. (c)~Imaginary (stable) eigenvalues as a function of~$\epsilon$.}
\label{case_1}
\end{figure*}
\begin{figure*}[tbph]
\centering
\subfigure[]{\includegraphics[width=0.35\textwidth]{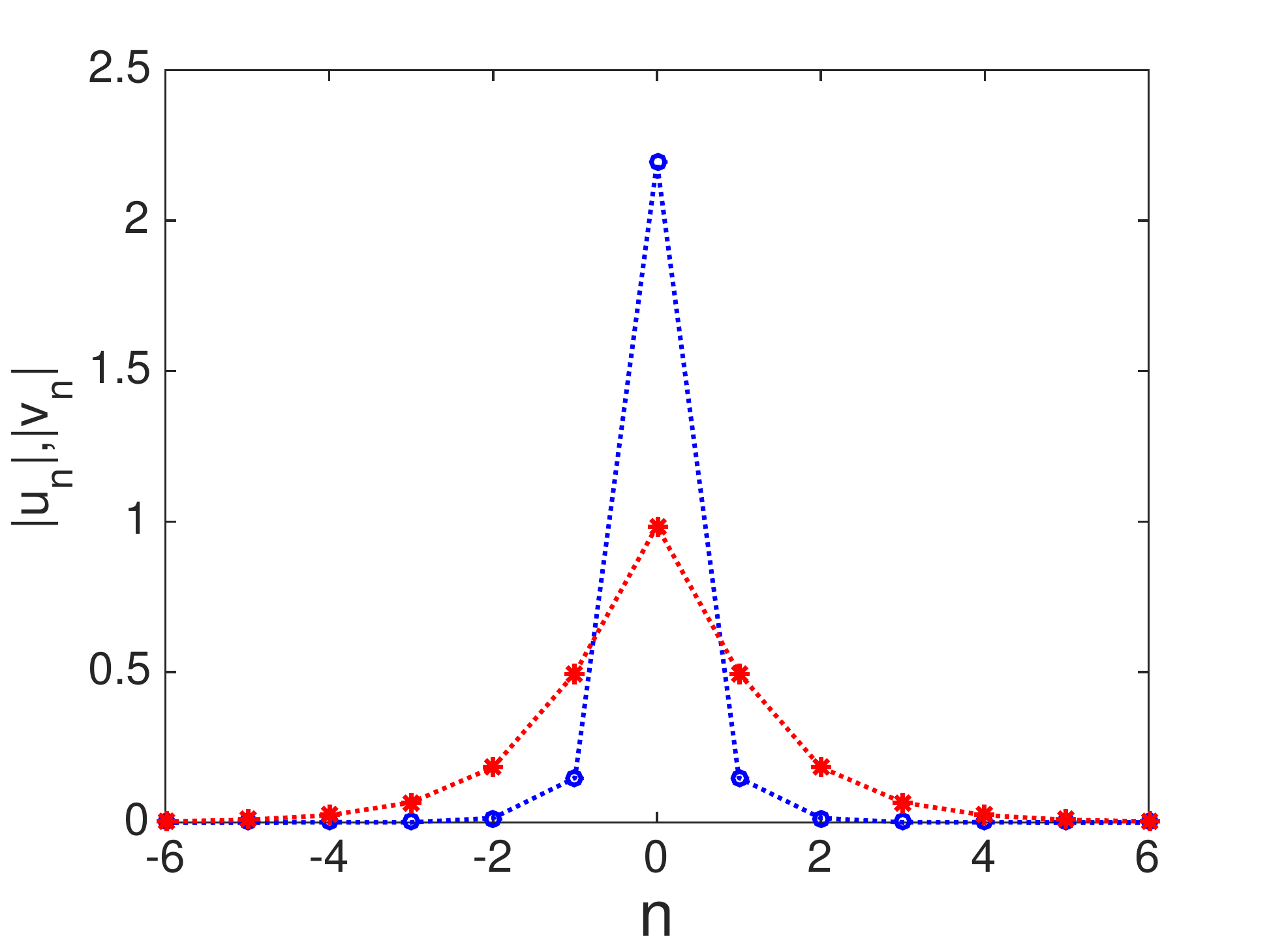}} \hspace{1cm}
\subfigure[]{\includegraphics[width=0.35\textwidth]{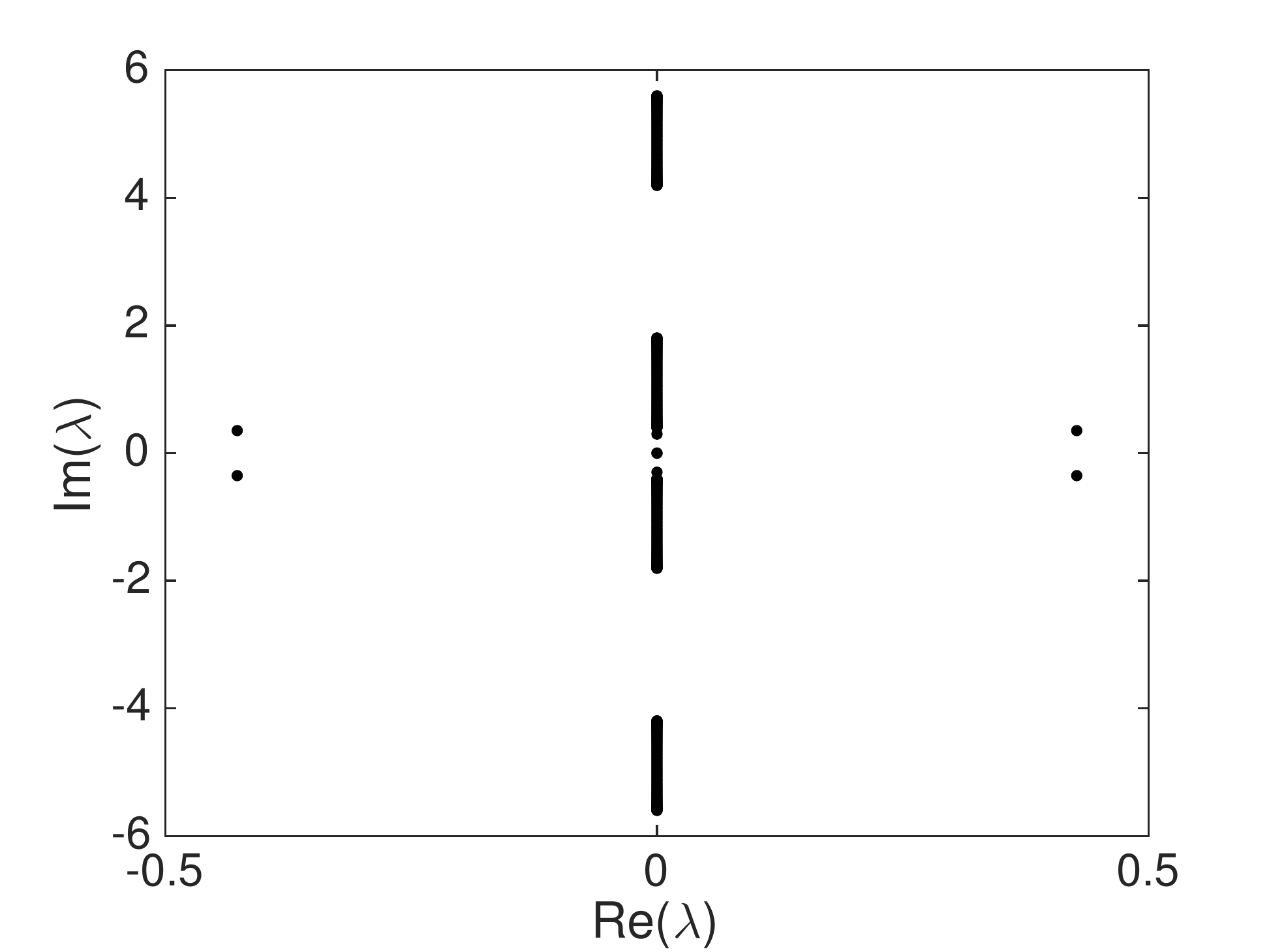}} \\
\subfigure[]{\includegraphics[width=0.35\textwidth]{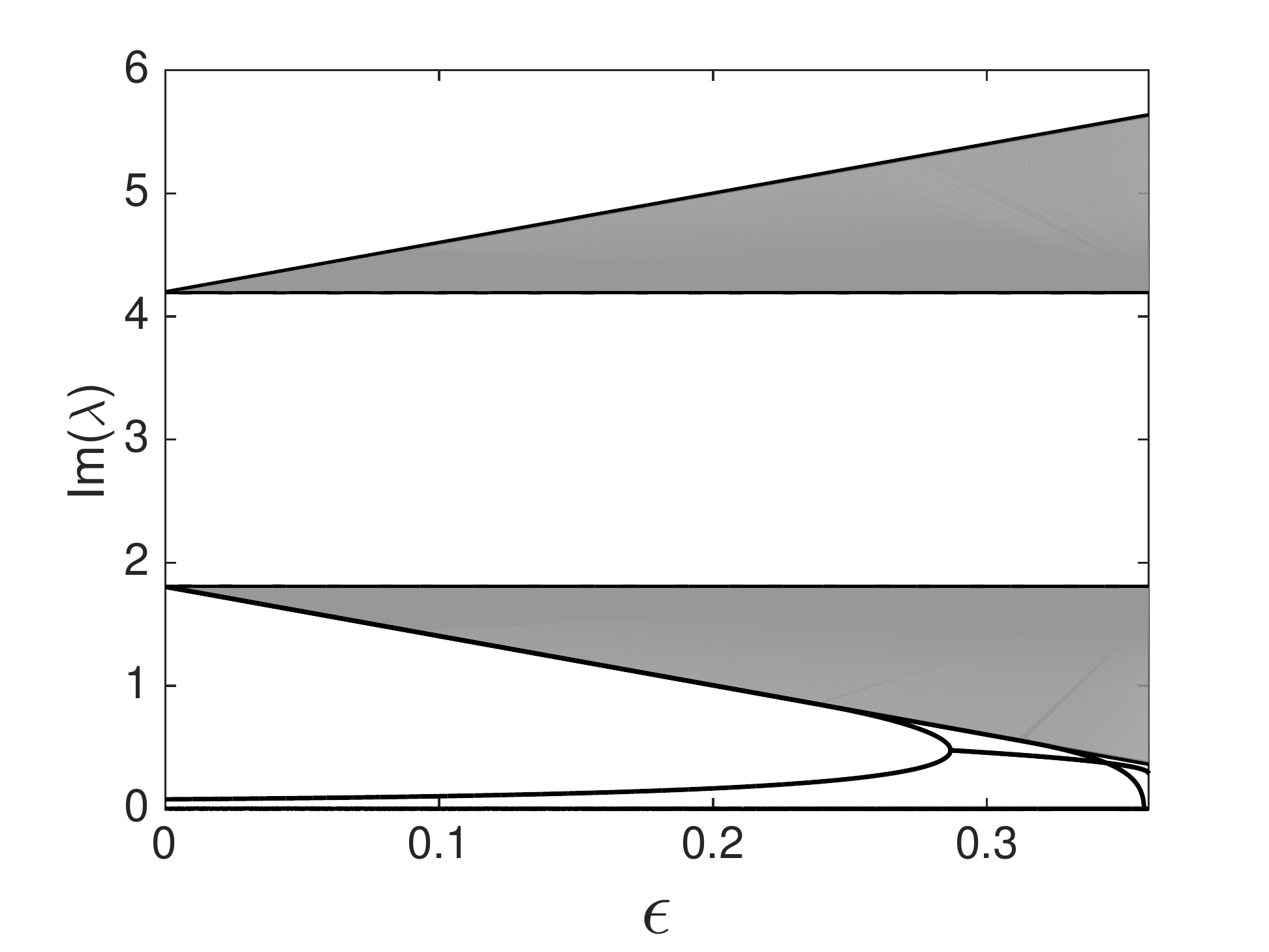}} 	 \hspace{1cm}
\subfigure[]{\includegraphics[width=0.35\textwidth]{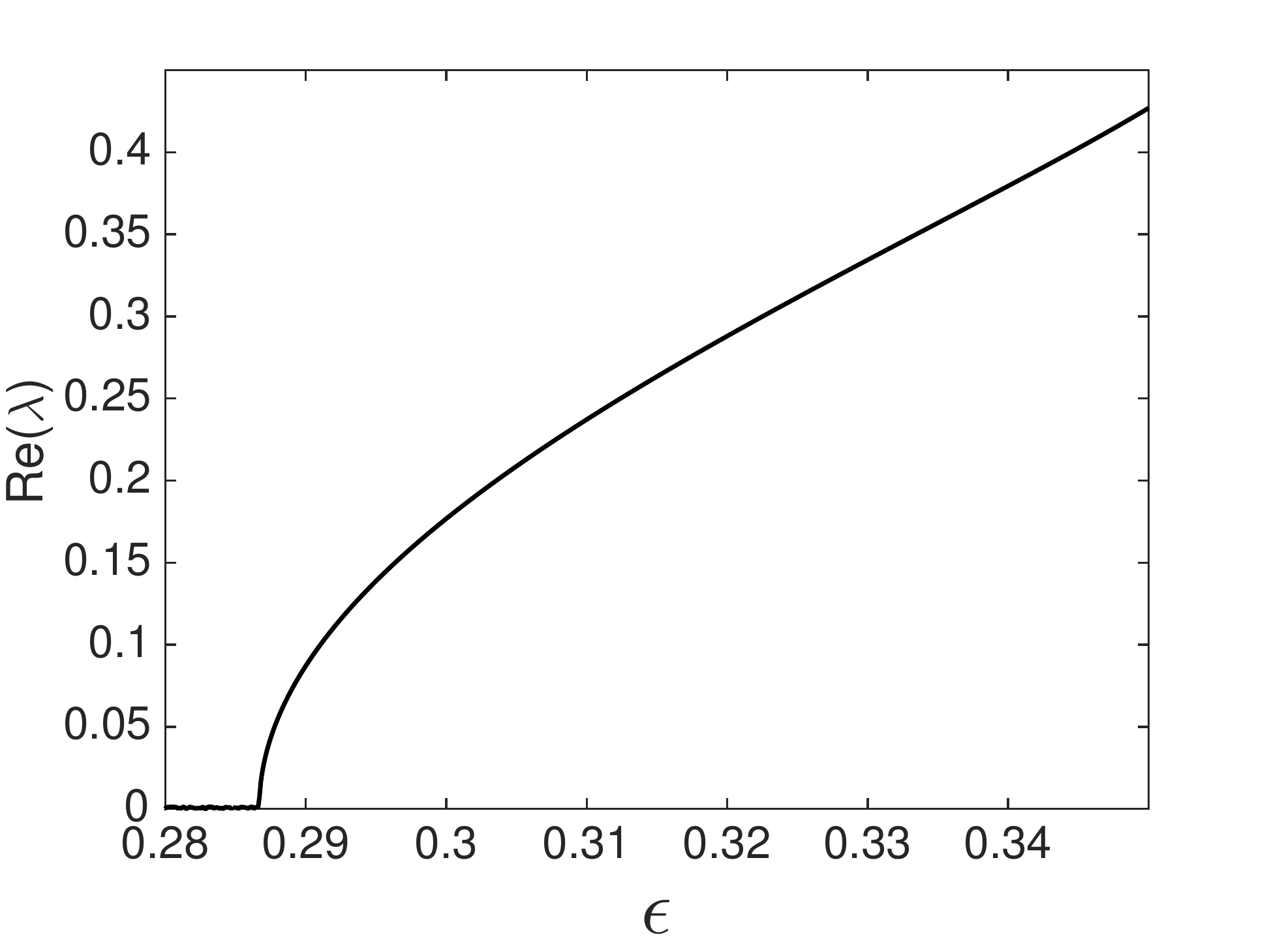}}
\caption{The same as Fig.~\ref{3rdcase1}, with $\epsilon = 0.35$, $K = -3$, $\gamma = 0.1$, and $q = 1.2$ but for \textit{in-phase} fundamental solitons corresponding to the analytical approximation based on Eqs.~\eqref{ds} and~\eqref{a0b0_12stnd}, with the ``$+$'' sign in the expressions for $\tilde{a}_0$ and $\tilde{b}_0$. In panel (a), the taller (blue) and shorter (red) curves correspond to $|u_n|$ and $|v_n|$, respectively. Panel (d) shows real (unstable) eigenvalues as a function of the intersite coupling $\epsilon$.}
\label{case_2}
\end{figure*}
\begin{figure*}[tbph]
\centering
\subfigure[]{\includegraphics[width=0.32\textwidth]{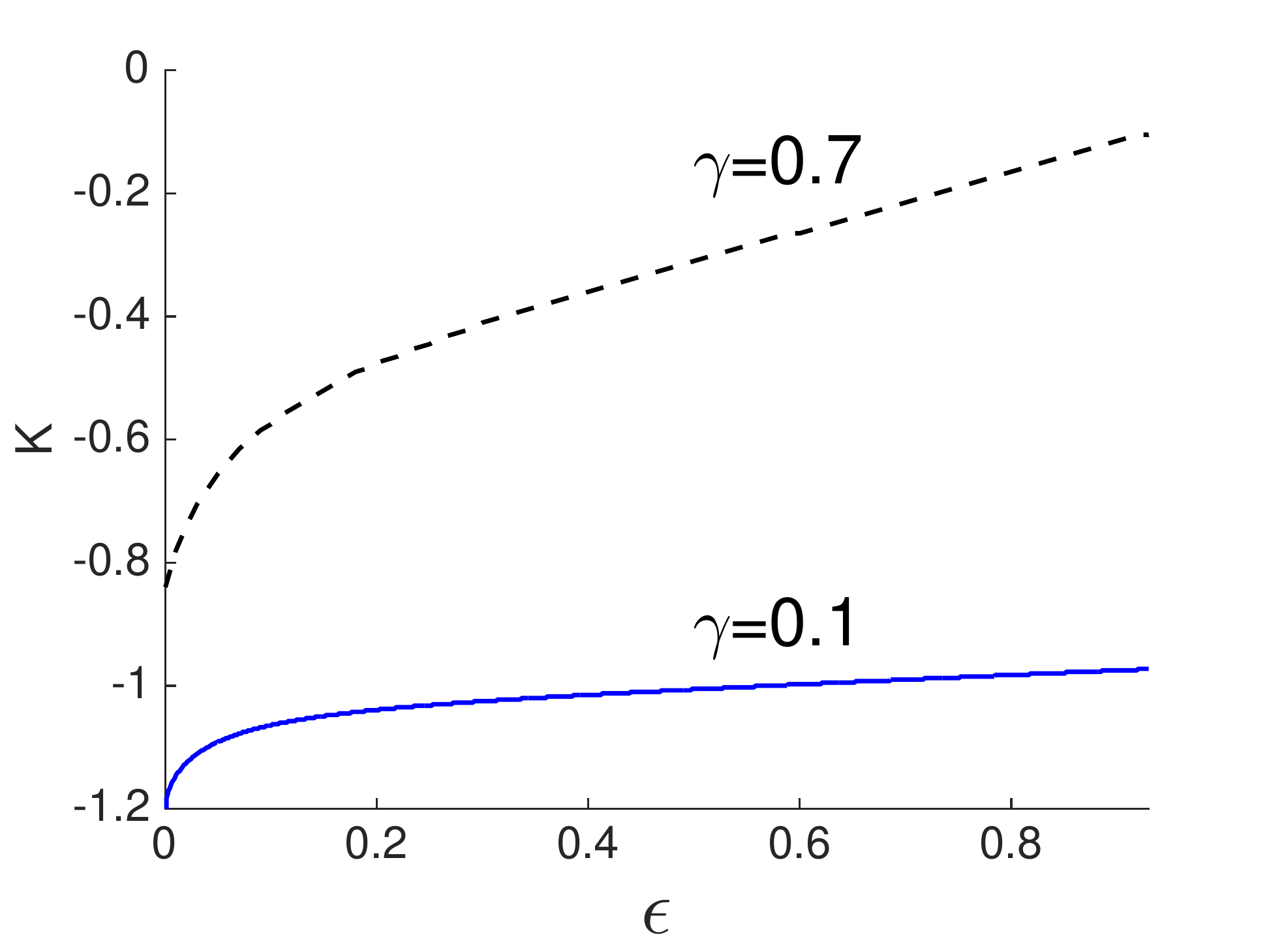}} 
\subfigure[]{\includegraphics[width=0.33\textwidth]{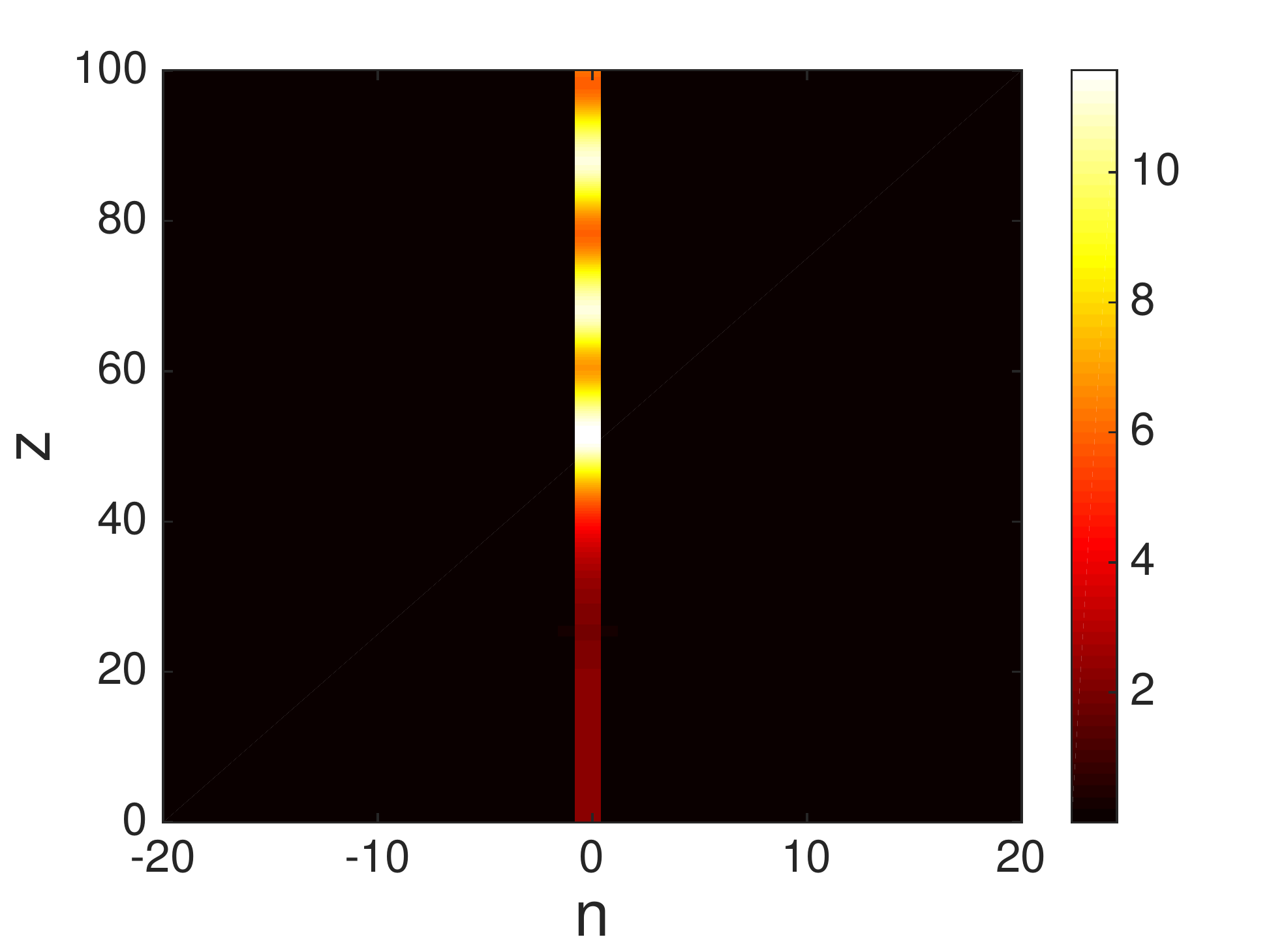}} 
\subfigure[]{\includegraphics[width=0.33\textwidth]{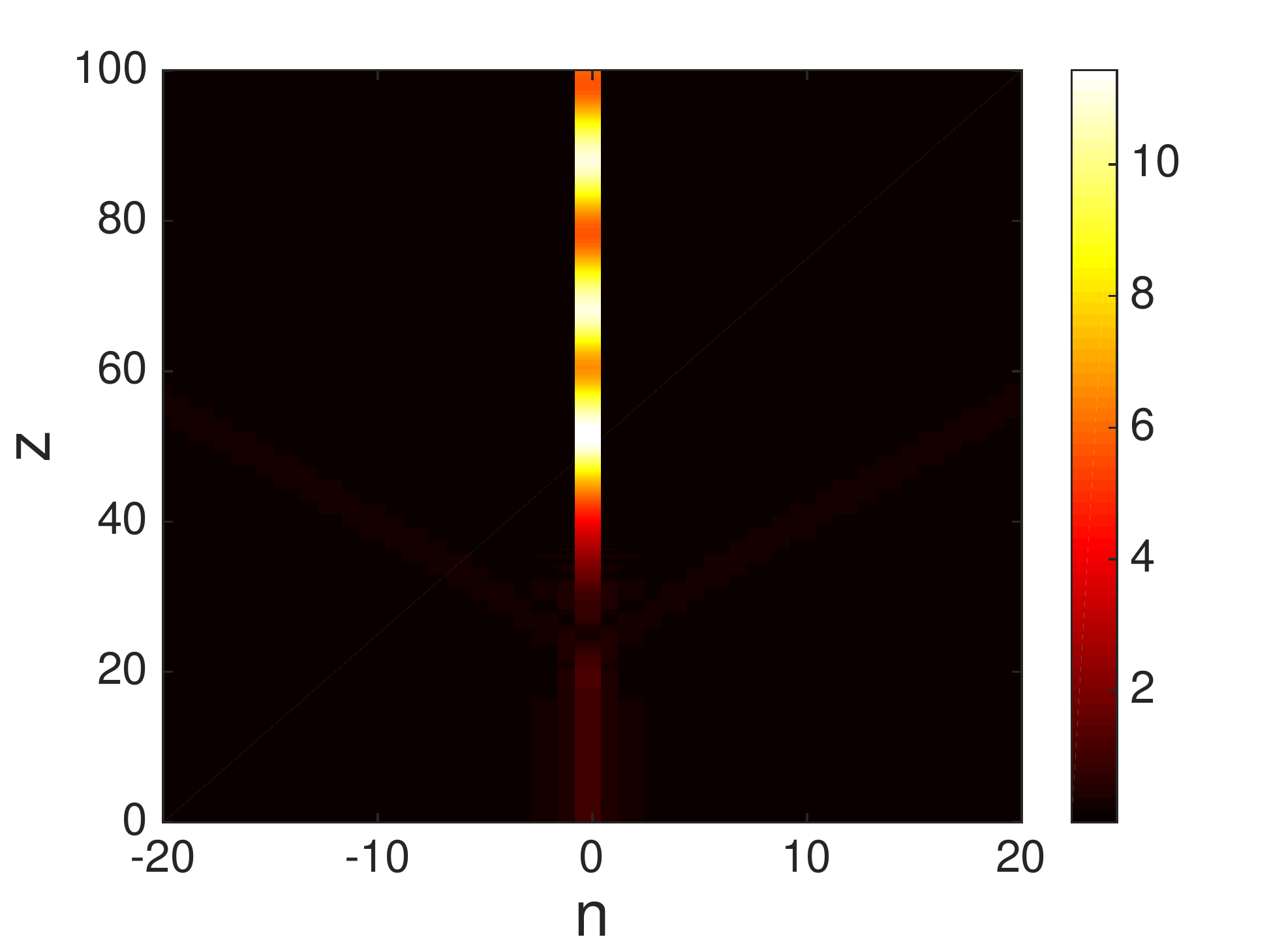}}
\caption{(a) The boundary of the instability region for in-phase discrete fundamental solitons (introduced in Fig.~\ref{case_2}), at two values of the intercore coupling constant $\gamma$. The solitons are unstable below the curves. (b, c) The evolution of an unstable in-phase discrete soliton for $\epsilon = 0.35$, whose stationary shape is displayed in Fig.~\ref{case_2}(a). Depicted in the panels (b) and (c) is the evolution of $|u_{n}|^2$ and $|v_{n}|^2$, respectively.}
\label{Dynamic_case_2}
\end{figure*}
\begin{figure*}[tbph]
\centering
\subfigure[]{\includegraphics[width=0.35\textwidth]{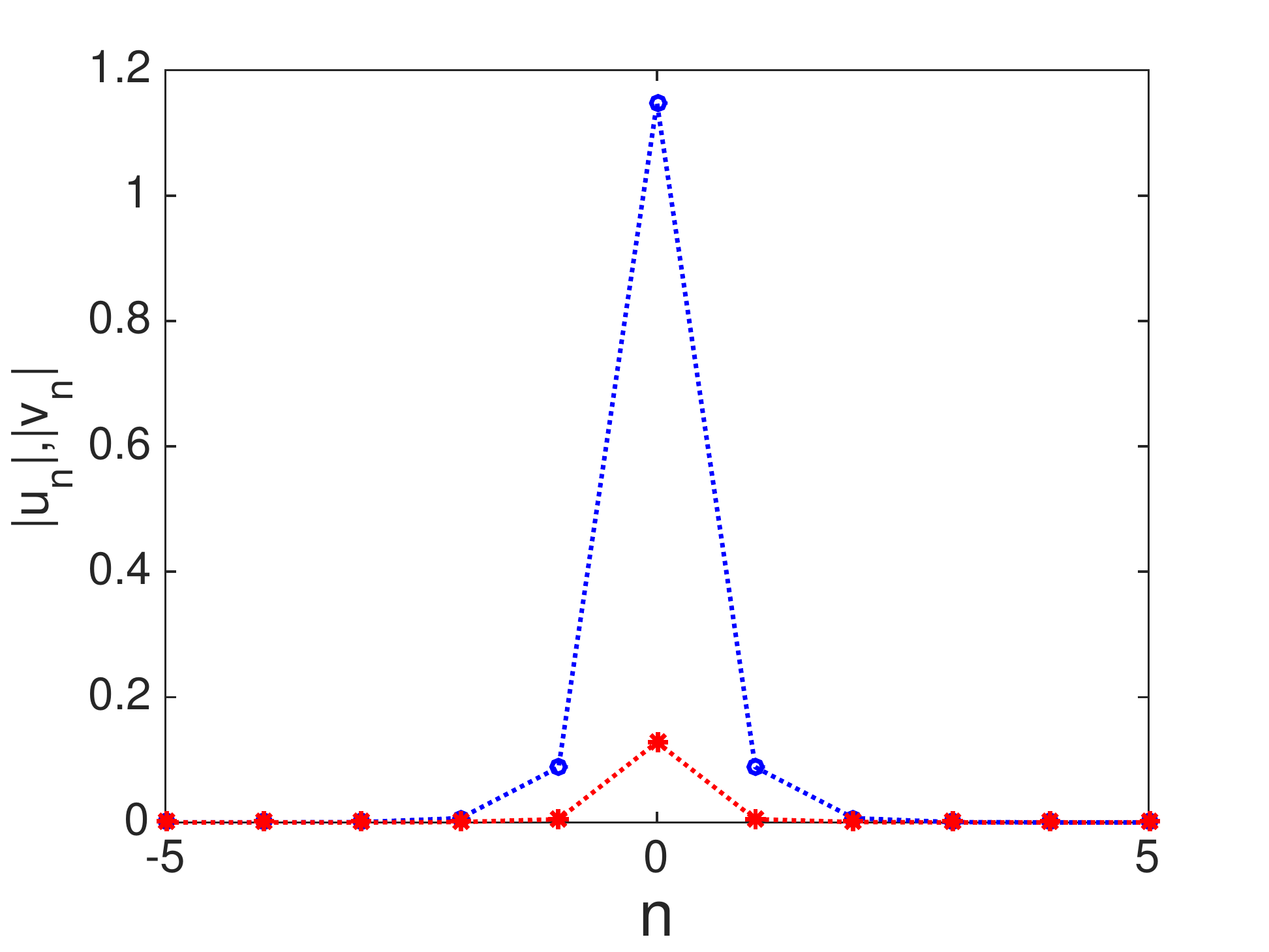}}  \hspace{1cm}
\subfigure[]{\includegraphics[width=0.35\textwidth]{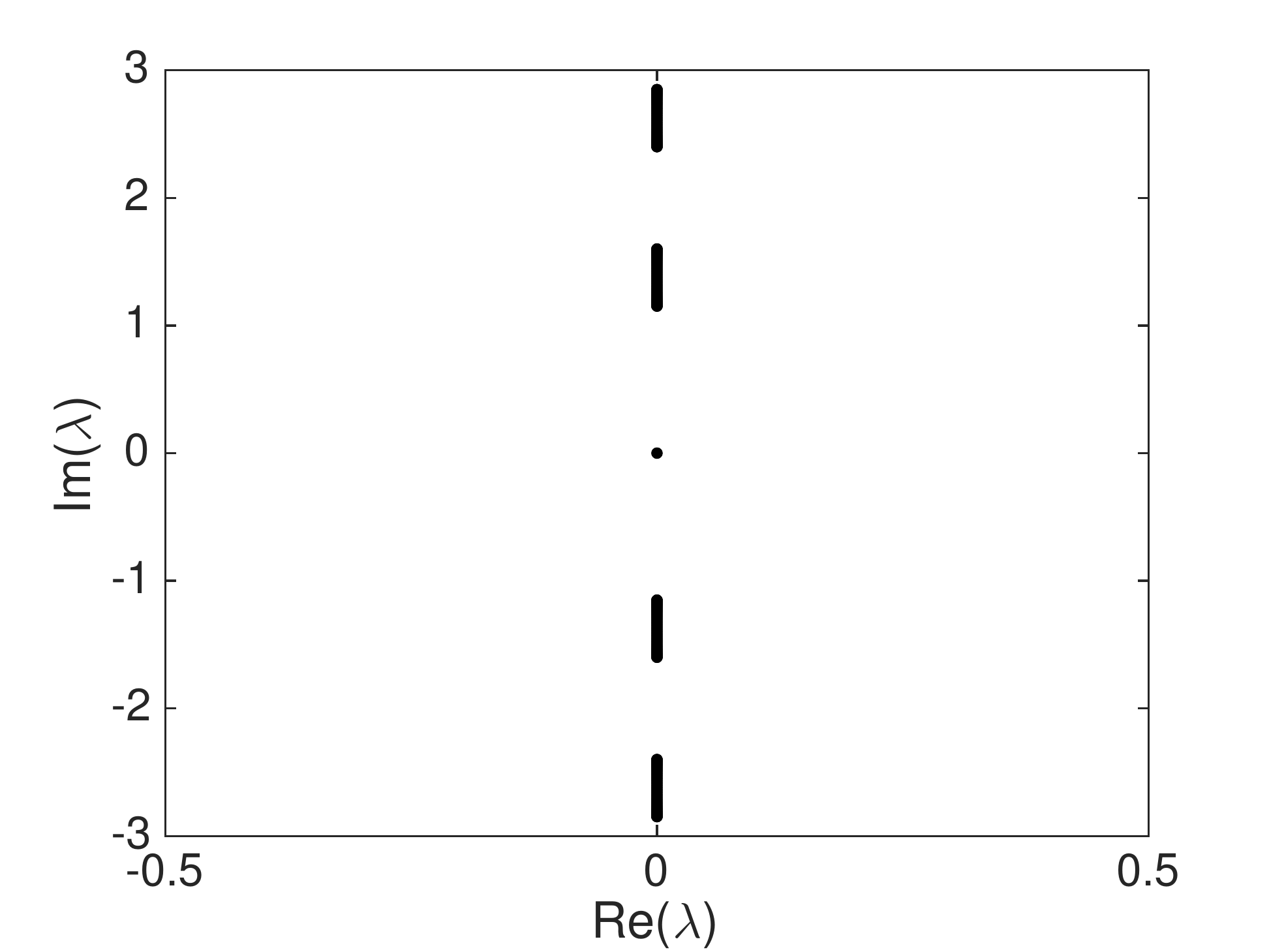}}  \\
\subfigure[]{\includegraphics[width=0.35\textwidth]{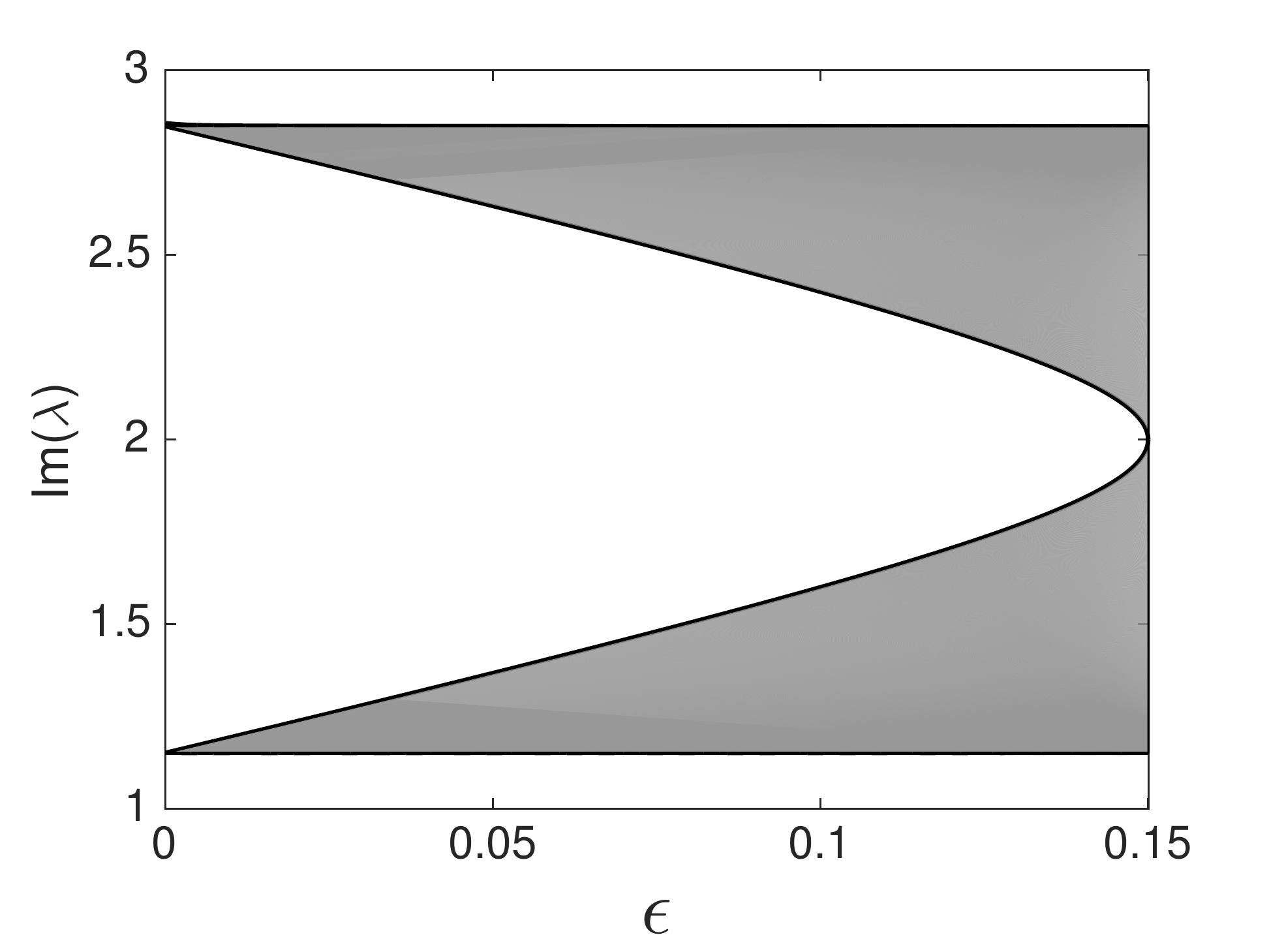}} \hspace{1cm}
\subfigure[]{\includegraphics[width=0.35\textwidth]{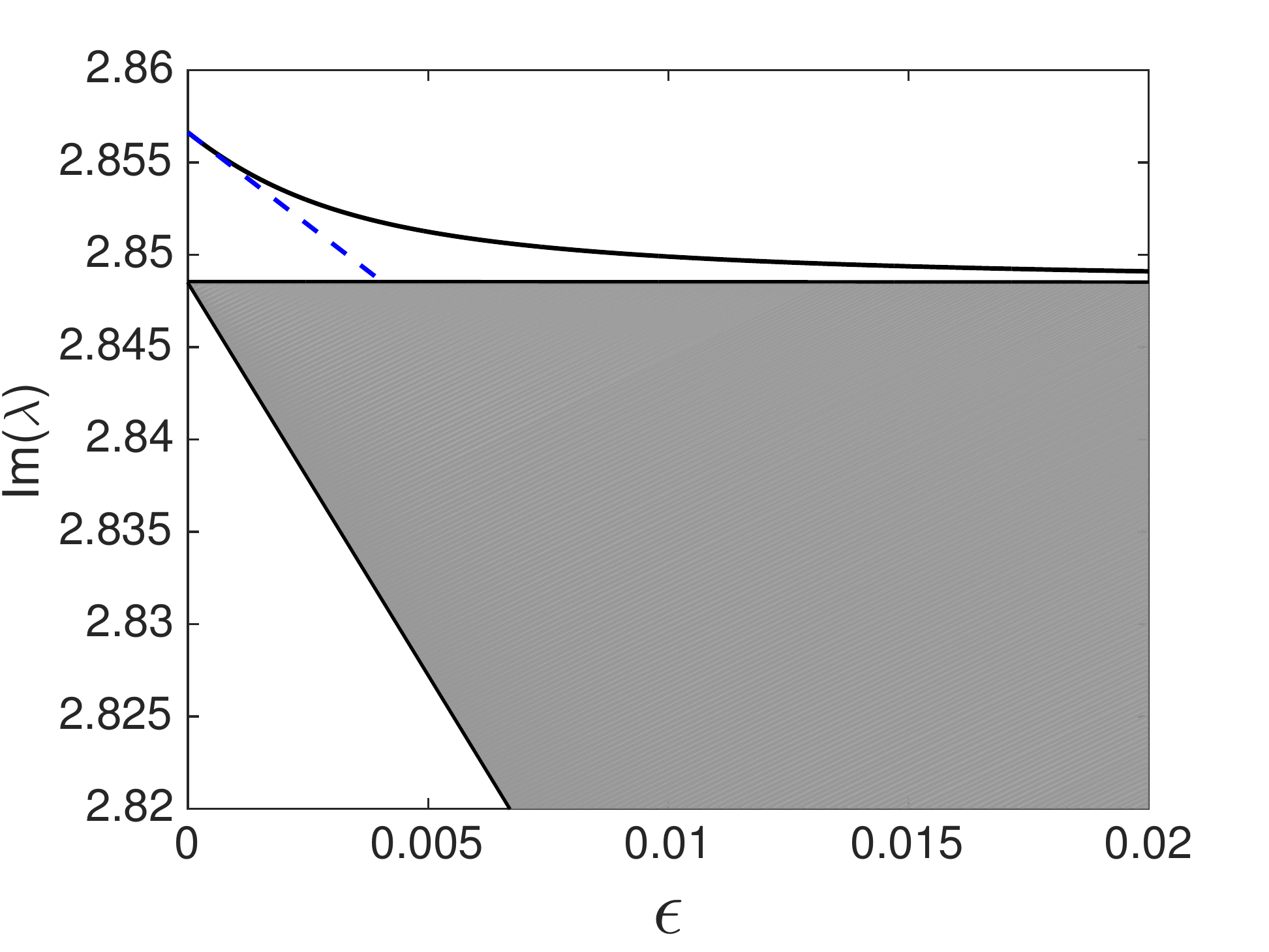}}
\caption{The same as Fig.~\ref{3rdcase1}, but for the discrete soliton initiated, in the approximate form, by Eqs.~\eqref{ds} and~\eqref{a0b0_3rd2}, and its stability spectrum for $\epsilon = 0.1$, $K = -2$, $\gamma = 0.3$, and $q = -0.9$. In panel (a), the taller (blue) and shorter (red) curves correspond to $|u_n|$ and $|v_n|$, respectively. The approximation for the separate eigenvalue is given by Eq.~\eqref{app11}, shown by the dashed line in panel (d).}
\label{case_3_q-1}
\end{figure*}

First, we have considered families of fundamental discrete solitons which are initiated, at small $\epsilon$, by the approximation based on Eqs.~\eqref{ds}, with $\tilde{a}_0$ and $\tilde{b}_0$ taken as per Eqs.~\eqref{a0b0_3rd}. As mentioned above, in the continuum limit, corresponding to $\epsilon \rightarrow \infty$, stable gap solitons exist under condition~\eqref{q>}, in the spectral gap defined by the first inequality in Eq.~\eqref{ww}~\cite{dana2015cp}. Our results demonstrate that, under the same conditions, there is a family of fundamental discrete solitons which carries over into its continuum-limit counterpart, which has been studied in detail in Ref.~\cite{dana2015cp}. In this case, the characteristics of the discrete solitons are quite similar to those found in the continuum limit; therefore in what follows we concentrate on solutions that \emph{do not exist} in the continuum limit, i.e., the respective families terminate before reaching the continuum limit. In all cases, this happens to fundamental discrete solitons belonging to semi-infinite band gaps, as these band gaps themselves are pushed out to infinity in the continuum limit.

In Fig.~\ref{3rdcase1}, we display numerical results for the fundamental-soliton family initiated by Eqs.~\eqref{ds}, with $\tilde{a}_0$ and $\tilde{b}_0$ again taken as per Eqs.~\eqref{a0b0_3rd}, while fixed (in this figure) $K = -3$ belongs to the semi-infinite band gap defined by the second inequality in Eq.~\eqref{ww}, rather than the first (finite) one. The analytical expression~\eqref{app1} for the separate eigenvalue is displayed too, showing reasonable proximity to its numerical counterpart. It is seen that these solutions are linearly stable. In this case, there is a critical (cutoff) value, $\epsilon_{\mathrm{cr}}$, of coupling constant $\epsilon$ at which the discrete-soliton family terminates. The cutoff can be readily explained, noting that in Fig.~\ref{3rdcase1} we choose $q > \gamma$, i.e., the second inequality in Eq.~\eqref{ww} holds, for given $K$, in the interval \\ \vspace*{-0.75cm}
\begin{align}
&-\epsilon_{\mathrm{cr}}^{+} < \epsilon < \epsilon_{\mathrm{cr}}^{-},			\label{interval} \\
\epsilon_{\mathrm{cr}}^{\pm} =& \frac{1}{4} \left(\sqrt{K^2 + \gamma^2} \pm q \right),  \label{cr2}
\end{align}
i.e., $-1.05 < \epsilon < 0.45$, in the present case ($K = -3,~q = 1.2,~\gamma = 0.1$). The cutoff value $\epsilon = \epsilon_{\mathrm{cr}}^{-} > 0$ in Fig.~\ref{3rdcase1} corresponds to the situation when the lower branch of the continuous spectrum [see panel (c)] touches the horizontal axis, signaling the onset of delocalization of the discrete soliton.

In Fig.~\ref{Dynamic_case_3}, we plot a typical example of the evolution of a discrete soliton \emph{past} the critical point; i.e., we use the discrete soliton, found at $0 < \epsilon_{\mathrm{cr}} - \epsilon \ll \epsilon_{\mathrm{cr}}$, as the input for direct simulations on the other side of the point, at $0 < \epsilon - \epsilon_{\mathrm{cr}} \ll \epsilon_{\mathrm{cr}}$. The simulations exhibit ``breathing'' dynamics, with a gradually decaying breathing amplitude of the second field, as seen in Fig.~\ref{Dynamic_case_3}(b). The decay is caused by an emission of radiation (linear waves) from the pulsating soliton. Thus, it indeed suffers the delocalization, gradually decaying via the radiation loss.

Next, we consider the family of discrete solitons which is initiated, in the analytical approximation, by Eqs.~\eqref{ds}, with $\tilde{a}_0$ and $\tilde{b}_0$ taken as per Eqs.~\eqref{a0b0_4thc}, assuming $K < -1$. This family also belongs to the semi-infinite gap, defined by the second inequality in Eq.~\eqref{ww} and by Eq.~\eqref{interval}. The solution profile and its stability are displayed in Fig.~\ref{4thcase1}. The approximation~\eqref{app2} is also presented, again showing reasonable agreement with the numerical findings. This branch of the discrete solitons again ceases to exist at $\epsilon > \epsilon_{\mathrm{cr}}$, when fixed $K$ leaves the semi-infinite gap.

To complete the analysis of the system with the positive phase-velocity mismatch, $q > 0$, we consider discrete solitons originating from the analytical approximation~\eqref{ds} with $\tilde{a}_0$ and $\tilde{b}_0$ given by Eqs.~\eqref{a0b0_12stnd}, which again requires $K < -1$ for its existence. Due to the ``$\pm$'' sign in Eqs.~\eqref{a0b0_12stnd}, there are two types of the solutions that we refer to as the in-phase and out-of-phase discrete solitons, which correspond, respectively, to identical and opposite signs of the two components, while both species are shaped as fundamental solitons.
\begin{figure*}[tbph]
\centering
\subfigure[]{\includegraphics[width=0.35\textwidth]{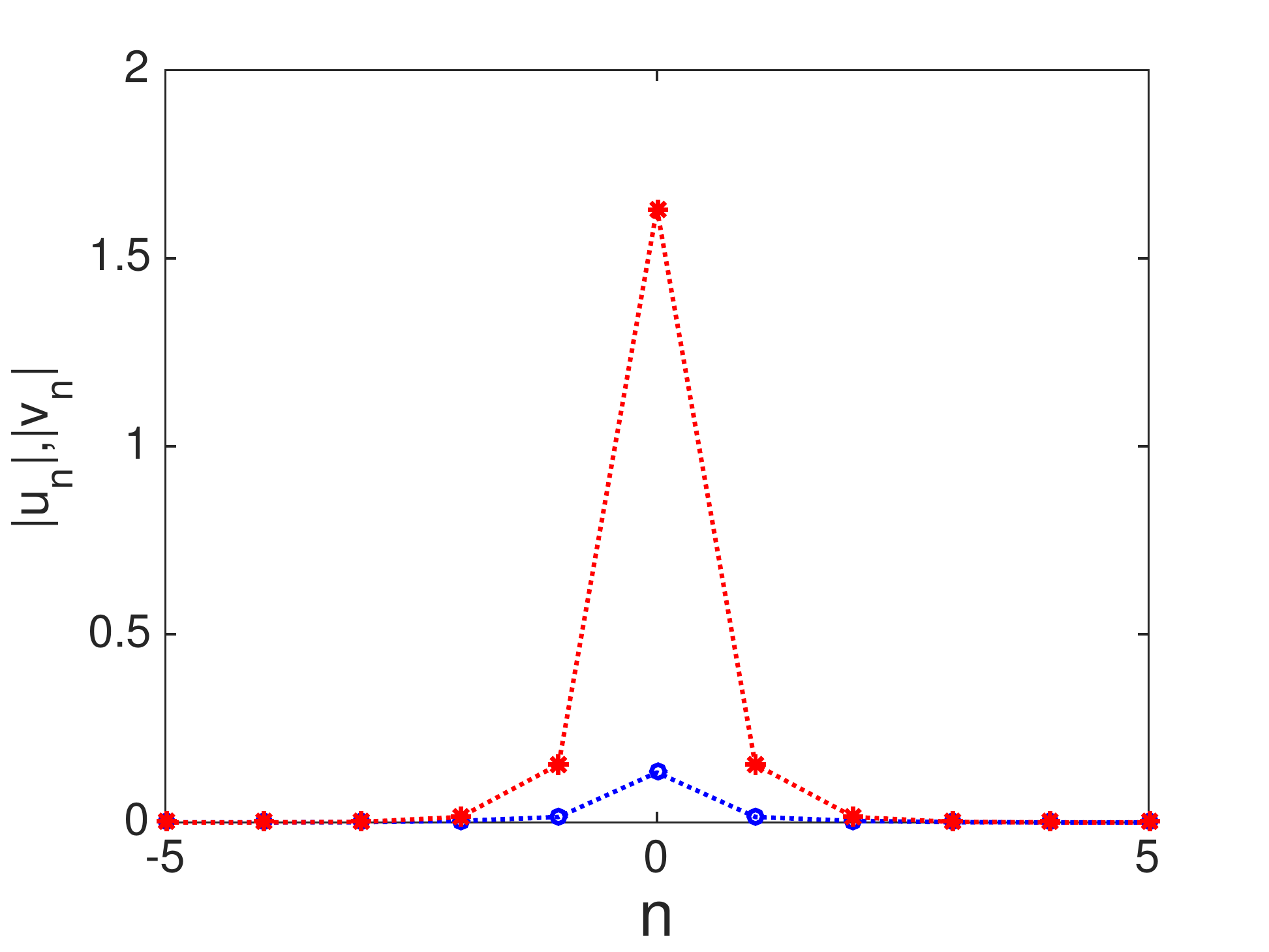}} \hspace{1cm}
\subfigure[]{\includegraphics[width=0.35\textwidth]{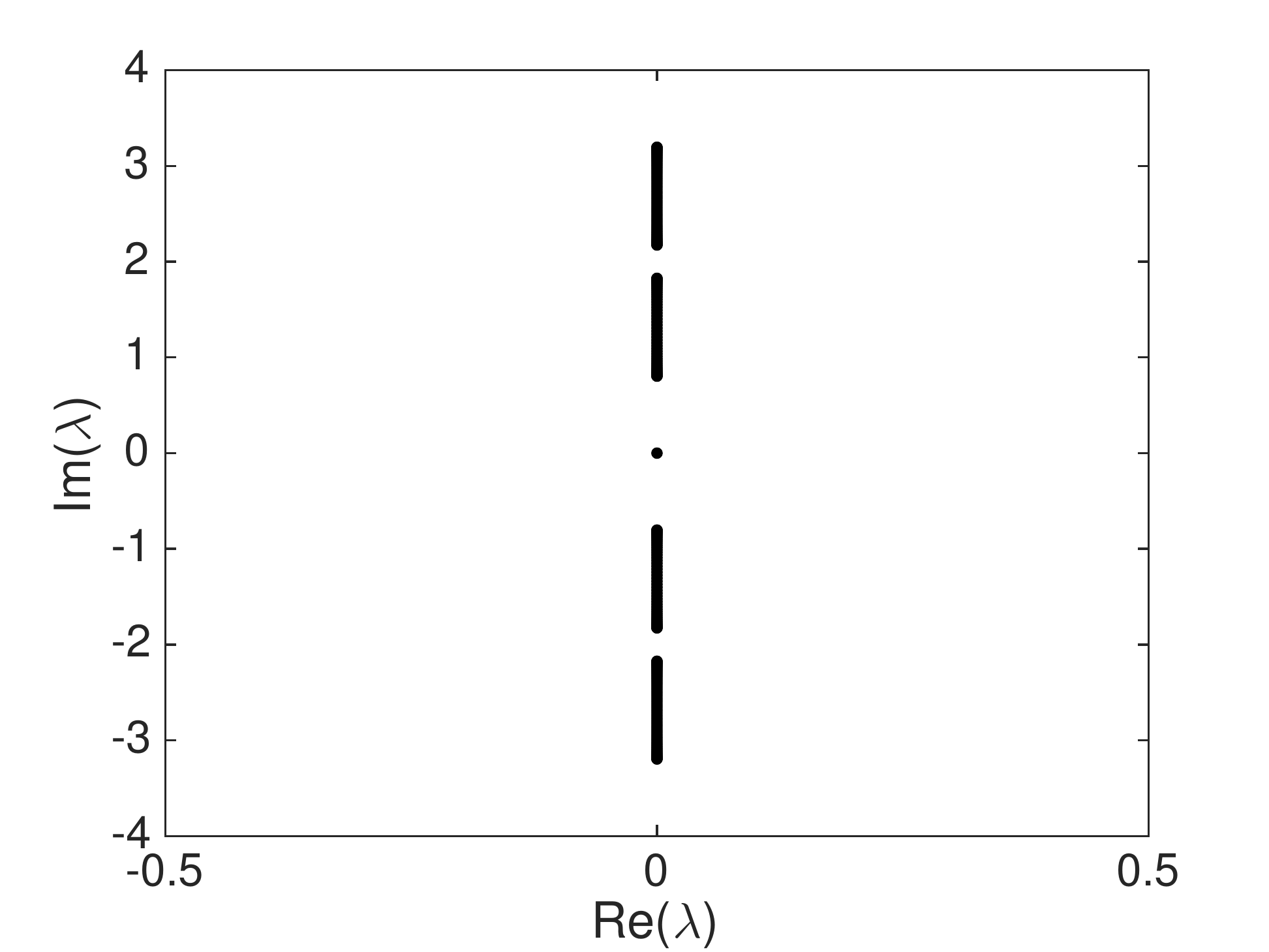}}   \\
\subfigure[]{\includegraphics[width=0.35\textwidth]{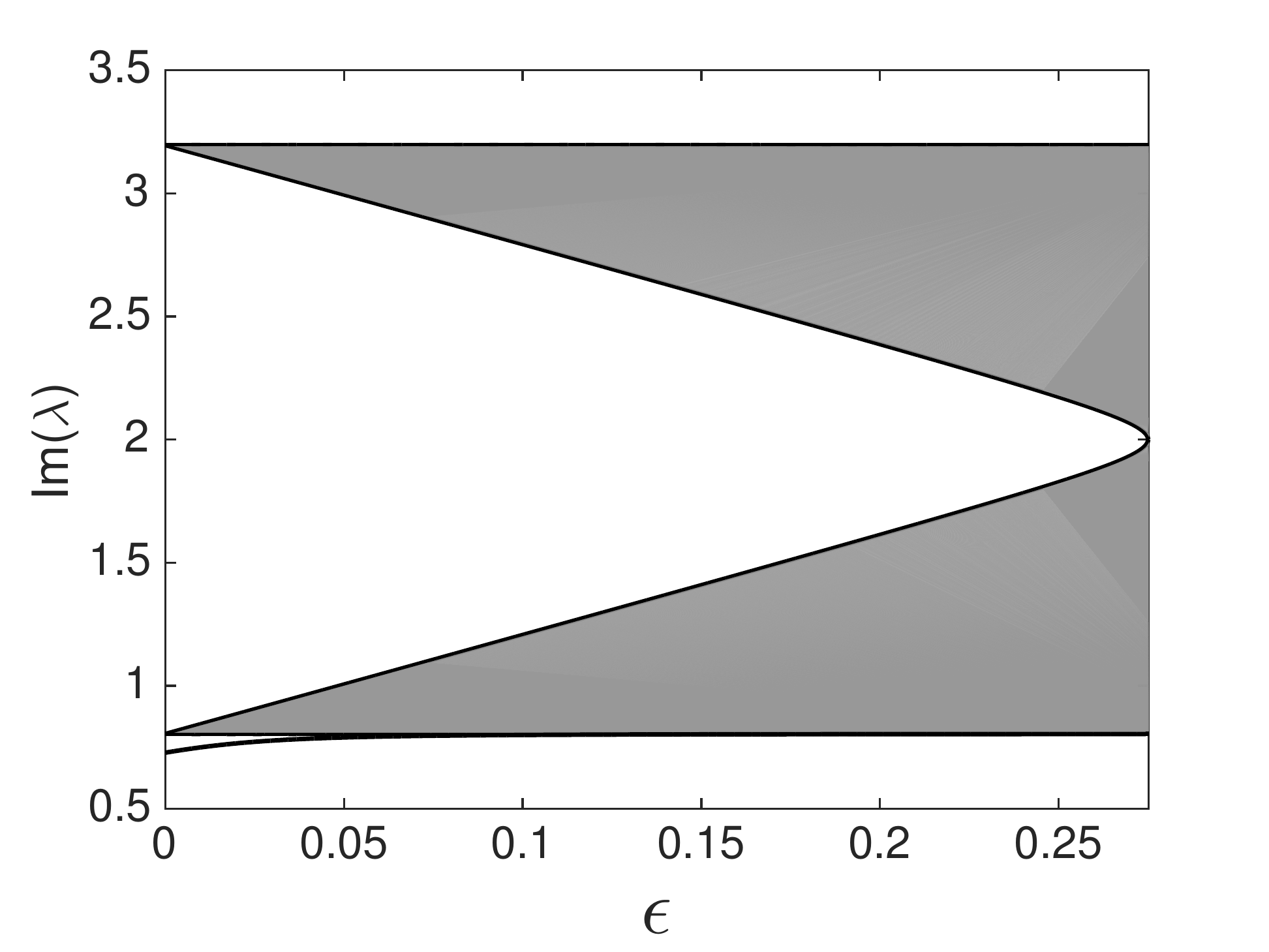}} \hspace{1cm}%
\subfigure[]{\includegraphics[width=0.35\textwidth]{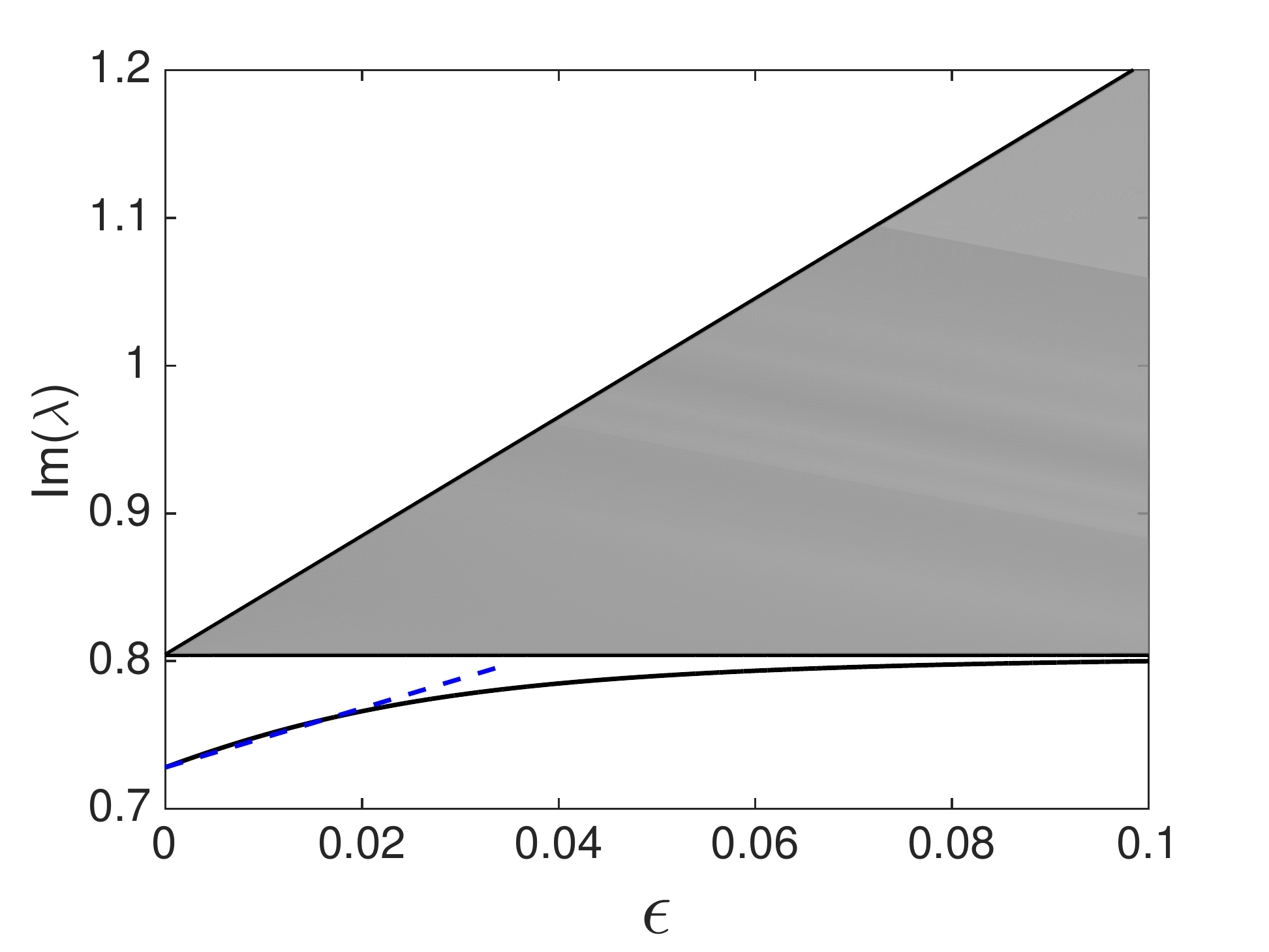}} 
\caption{The same as Fig.~\ref{3rdcase1}, but for the discrete soliton initiated, in the approximate form, by Eqs.~\eqref{ds} and~\eqref{a0b0_4th2}, and its stability for $\epsilon = 0.1$, $K = -2$, $\gamma = 0.1$, and $q = -1.2$. In panel (a), the shorter (blue) and taller (red) curves correspond to $|u_n|$ and $|v_n|$, respectively. The approximation for the separate eigenvalue is given by Eq.~\eqref{app22}, shown by the dashed line in panel (d).}
\label{casetm}
\end{figure*}

The profile and stability of the out-of-phase solitons are shown in Fig.~\ref{case_1}, where one can see that the solitons are again \emph{stable} in their \emph{entire} existence region. For the chosen parameters, $K = -3$, $\gamma = 0.1$, and $q = 1.2$, we obtain from Eqs.~\eqref{interval} and~\eqref{cr2} that the semi-infinite gap is bounded by $\epsilon_{\mathrm{cr}}^{-} = 0.4525$. This agrees with the numerical results in Fig.~\ref{case_1}, where the soliton family can only be computed up to the critical coupling, which is located beyond the frame of Fig.~\ref{case_1}(c).

Further, we depict the same for the in-phase solitons in Fig.~\ref{case_2}. Different from their out-of-phase counterpart, these species of the discrete fundamental solitons become unstable beyond a critical point, which is found inside of its existence interval. The instability is caused by a collision of two eigenvalues on the imaginary axis (where one of them bifurcates from the continuous spectrum), thus creating a quartet of complex eigenvalues, i.e., giving rise to oscillatory instability. This is a known generic scenario of the onset of instability of discrete solitons (cf. Refs.~\cite{MK,kiri16}). The stability region, as well as typical evolution initiated by the instability, is shown in Fig.~\ref{Dynamic_case_2}. It is clearly seen that the amplitude of the unstable solution increases with oscillations, indicating an eventual blow-up (recall that we are dealing with a nonconservative system, where such an outcome is possible).

We have also considered the case of $q < 0$, i.e. negative phase-velocity mismatch in Eqs.~\eqref{1} and~\eqref{2}. In this case, the discrete fundamental solitons belong to the semi-infinite gap defined by the second inequality in Eqs.~\eqref{ww2} and~\eqref{q<0}. For fixed $q$ and $\gamma$, the existence range of the solitons cannot be extended towards the continuum limit ($\epsilon \rightarrow \infty$), as Eq.~\eqref{q<0} imposes the limitation,
\begin{equation}
\epsilon < \frac{1}{4} \left(|q| - \gamma \right).  			\label{1/4}
\end{equation}
In Figs.~\ref{case_3_q-1} and~\ref{casetm} we display the discrete solitons which are initiated by the analytical approximation based on Eqs.~\eqref{ds} with $\tilde{a}_0$ and $\tilde{b}_0$ taken as per Eqs.~\eqref{a0b0_3rd2} and~\eqref{a0b0_4th2}, respectively. We also plot the analytical approximation for the separate eigenvalue given by Eqs.~\eqref{app11} and \eqref{app22}, where good agreement is again observed. In panel (c) of both Figs.~\ref{case_3_q-1} and~\ref{casetm}, the critical value of the coupling constant $\epsilon$, above which condition~\eqref{1/4} does not hold, corresponds to the situation when the two branches of the continuous spectrum merge. In this case, we do not display numerical results for discrete solitons initiated by the analytical approximation based on Eqs.~\eqref{ds}, with $\tilde{a}_0$ and $\tilde{b}_0$ taken as per Eqs.~\eqref{a0b0_12stnd2}, because the respective results for stable solutions are quite similar to those displayed in Figs.~\ref{case_3_q-1} and~\ref{casetm}.

\section{Conclusion}						\label{sec5}

In this work, we have presented a model of the dual-core optical waveguiding array, which may be used to emulate the $\mathcal{CP}$-symmetry in the discrete system. Necessary ingredients of the system are opposite signs of the discrete diffraction in the two parallel arrays (cores), which may be implemented by means of the diffraction-management technique, and the active coupling between the arrays, which accounts for the gain and loss in the system, the stability of the zero state being provided by a sufficiently large phase-velocity mismatch between the parallel arrays. The analytical results, obtained by means of the extension from the anticontinuum limit, and the numerical findings show the existence of several families of discrete fundamental solitons in the system. Unlike the continuum limit of the present setting, considered in Ref.~\cite{dana2015cp}, which maintains a single family of gap solitons, the discrete system supports different types of self-trapped modes, with the propagation constant falling into semi-infinite gaps of the corresponding linear spectrum. Most soliton families are stable, except for one, which develops the oscillatory instability past the internal stability boundary, as shown in Figs.~\ref{case_2} and~\ref{Dynamic_case_2}. 		

The family populating the finite band gap extends to the continuum limit, carrying over into the abovementioned stable gap solitons, while other branches terminate by hitting the edge of the semi-infinite gaps and suffering delocalization in this case. Species of higher-order discrete solitons, which may be stable but disappear or suffer destabilization in the continuum limit, are known in conservative systems, such as the 1D twisted states and 2D localized vortices in the discrete NLSEs. Here, continuous soliton families which exist solely in the discrete setting are reported in the non-Hermitian system. On the contrary to the abovementioned findings in conservative models, in the system that we currently study, these are families of fundamental solitons, which feature a noteworthy property of being completely stable (with the exception of one partially stable branch) in their existence regions.

Another essential difference from the previously studied systems is the fact that the discrete soliton families reported in this work are controlled not by the single parameter, viz., the effective strength of the intersite coupling ($\epsilon$, in the present notation), but also by the phase-velocity mismatch, $q$, and intercore coupling constant in the active host medium, $\gamma$. This conclusion suggests significant implications for the experimental creation of such solitons, because $\gamma$ can be readily adjusted by varying the gain which maintains the active host medium (e.g., this may be the power of the second-harmonic pump which realizes the scheme in terms of the mismatched three-wave system~\cite{dana2015cp}).

A natural extension of the present work may be the search for higher-order discrete solitons, such as twisted (dipole) and multipole states, in addition to the fundamental solitons presented here. A challenging direction for the further work is investigation of the 2D version of the system, realized as a square-shaped network of $\mathcal{CP}$-symmetric coupled waveguiding arrays. In particular, it may be interesting to construct stable 2D solitons with embedded vorticity.

\vspace*{-0.25cm}
\section*{Acknowledgement}

R.K. gratefully acknowledges financial support from the Indonesia Endowment Fund for Education ({\slshape Lembaga Pengelolaan Dana Pendidikan, LPDP}) through Grant No. S-34/LPDP.3/2017.~B.A.M. is supported, in part, by the Joint Program in Physics between the NSF and the Binational (US-Israel) Science Foundation through Project No.~2015616, and by the Israel Science Foundation through Grant No.~12876/17.~N.K. acknowledges supports from the SKKU Samsung Intramural Research Fund No.~2016-1299-000 and the National Research Foundation of Korea through Grant No.~NRF-2017-R1C1B5-017743.~B.A.M. and N.K. appreciate the hospitality of the Department of Mathematical Sciences at the University of Essex (Colchester, UK).


\end{document}